%% file: main.tex
\documentclass{article}

\usepackage[margin=1in]{geometry}
\usepackage{amsmath}
\usepackage[nopatch]{microtype}
\usepackage{graphicx}
\usepackage{rotating}
\usepackage{booktabs}
\usepackage{amssymb}
\usepackage{bm}
\usepackage{mathrsfs}
\usepackage{comment}
\usepackage{centernot}
 \usepackage{enumitem}

\newtheorem{assumption}{Assumption}

\usepackage{enumitem}
\usepackage{natbib}

\date{August 28, 2025}
\title{The Power of Prognosis: Improving Covariate Balance Tests with Outcome Information}

\author{Clara Bicalho\footnote{Tinker Postdoctoral Fellow, Stanford University}, Adam Bouyamourn\footnote{Postdoctoral fellow, Department of Politics, Princeton University}, and Thad Dunning\footnote{Professor of Political Science, Department of Political Science, University of California, Berkeley.}\thanks{Corresponding author: thad.dunning@berkeley.edu}}
\begin{document}

%%% DO NOT REMOVE THESE LINES. For automatic word count.
%TC:ignore

\maketitle

\begin{abstract}
Scholars frequently use covariate balance tests to test the validity of natural experiments and related designs. Unfortunately, when measured covariates are unrelated to potential outcomes, balance is uninformative about key identification conditions. We show that balance tests can then lead to erroneous conclusions. To build stronger tests, researchers should identify covariates that are jointly predictive of potential outcomes; formally measure and report covariate prognosis; and prioritize the most individually informative variables in tests. Building on prior research on ``prognostic scores," we develop bootstrap balance tests that upweight covariates associated with the outcome. We adapt this approach for regression-discontinuity designs and use simulations to compare weighting methods based on linear regression and more flexible methods, including machine learning. The results show how prognosis weighting can avoid both false negatives and false positives. To illustrate key points, we study empirical examples from a sample of published studies, including an important debate over close elections. 
\end{abstract}

%%% DO NOT REMOVE THIS LINE. For automatic word count.
%TC:endignore

\section{Introduction}

Methodologists urge researchers to test observable implications of assumptions that aid causal inference. In natural experiments and related designs, researchers often report covariate balance tests. The logic appears straightforward: if a coin flip had determined treatment assignment, pre-treatment covariates or ``placebo outcomes" would have the same distribution, in expectation, in treatment and control groups (\citealt{EggersTunonDafoe2023}, \citealt{CaugheyDafoeSeawright2017}). A statistically insignificant association between treatment and covariates is consistent with random assignment---an important advantage, if true, for making inferences about causation---while a significant association may suggest a flaw in the design. 

Unfortunately, these widely used tests may shed no light on key identification conditions. Researchers testing the validity of an alleged natural experiment would like to know whether a treatment is assigned independently of potential outcomes---a condition sometimes called "as-if" random.\footnote{Potential outcomes are the outcomes that would be realized under counterfactual assignment to different treatments (\citealt{Neyman1923}, \citealt{Rubin1974}). \citet{Freedman_2009} and \citet{Dunning2012} discuss as-if random.} In many studies, however, none of the measured covariates used in the balance tests are prognostic, that is, associated with potential outcomes. In others, some covariates are prognostic but others are not. As we show in this article, balance tests based on irrelevant covariates unrelated to potential outcomes cannot tell us whether as-if random is plausibly met. A similar point applies to regression-discontinuity designs, in which analysts often seek to test the continuity of potential outcomes at a threshold determining treatment assignment, rather than as-if random.  The continuity of non-prognostic covariates is uninformative about the continuity of potential outcomes themselves.

Using a sample of experiments, natural experiments, and regression-discontinuity designs published in top political science journals, we demonstrate three features---and, we argue, problems---of existing balance tests. (1) Covariate prognosis is rarely or never measured. (2) The overall prognosis of the covariates used in balance tests varies across studies---and is often very weak. (3) Within studies, the prognosis of individual covariates also varies and is also often weak.

These problems undermine the ability of standard balance tests to assess identification conditions. In typical practice, researchers report lengthy covariate balance tables without considering which variables are actually predictive of outcomes.  However, sometimes imbalances occur on irrelevant covariates unrelated to potential outcomes. Other times, it is the prognostic covariates that are imbalanced.  Because prognosis is not measured or incorporated formally into tests, the varied informativeness of different covariates is not considered, and it is difficult to assess how meaningful are rejections for any individual covariate. Moreover, because researchers often present tests for different covariates separately, standard procedures also lead to  multiple testing problems as well as indeterminacy: there is no clear rule for rejecting an overall null hypothesis like as-if random. Finally, because researchers do not assess the joint prognosis of covariates, readers and reviewers cannot readily assess the overall power of the tests to falsify identification conditions.

We make several contributions in this article to addressing these common problems.  First, we show why covariate prognosis is important for balance tests. We demonstrate that tests using irrelevant, non-prognostic covariates can lead researchers falsely to reject as-if random when it is true or to fail to reject when it is false. 

Second, we show that researchers can increase the power and specificity of their tests by measuring the most jointly prognostic covariates possible---and then prioritizing the most informative covariates, among those they measure. Thus, we offer the following concrete advice for how researchers using balance tests can address problems (1)-(3) highlighted above:

\begin{enumerate}[label = (\arabic*)]
\item \textbf{Measure and report prognosis in balance tests}. We propose measures of covariate prognosis that help researchers and readers assess the informativeness of balance tests. Although potential outcomes are partially unobservable (\citealt{Holland_1986}), it is possible to assess, for instance, how well covariates predict potential outcomes under control using data from a control group sample. Such diagnostic measures are essential because prognosis is an empirical question. For example, while the pre-treatment values of outcome variables tend to be related to potential outcomes under control (\citealt{imbens_rubin_2015}: 483-4), such lagged outcomes may or may not be available to researchers; and, as we show, in some applications they are not in fact prognostic. 

\item \textbf{Maximize overall covariate informativeness}. Our results show that as the set of covariates used in balance tests become more prognostic, the tests become more powerful and specific. Thus, researchers should endeavor to collect data on the most jointly prognostic covariates possible. Theoretical and substantive knowledge can guide the identification of covariates that are likely associated with potential outcomes in a given context. Where feasible, researchers should include the lagged dependent variable as a covariate. Formalizing the reporting of measures of prognosis as part of the publication process, per (1), can aid assessment of informativeness and heighten researchers' incentives to gather data on the most predictive covariates possible. 

\item \textbf{Prioritize prognostic variables in the tests}. Finally, we utilize balance tests that prioritize---among a set of measured covariates---the individual variables most associated with potential outcomes. These tests upweight prognostic and downweight non-prognostic covariates to form a single informativeness-weighted test statistic. By combining information on prognosis across covariates into one omnibus test statistic, these tests confront problems of indeterminacy in covariate-by-covariate tests and can increase power and specificity.
\end{enumerate}
For (3), we build on research in statistics and epidemiology on the use of "prognostic scores" to estimate treatment effects in observational studies (especially \citealt{Hansen_2008}; also \citealt{RubinThomas_2000}, \citealt{Stuart_2013}, \citealt{Leacy_2014}, and \citealt{Wainstein_2022}). These methods have been very little used in the social sciences: only a handful of citations appear in social science journals, and prognostic scores are nearly absent from applied work in political science. 

We extend this previous work to construct covariate balance tests appropriate for different identification conditions. The key test statistic is the difference in the average fitted, covariate-adjusted potential outcome under control across treatment and control groups, as in \citet{Hansen_2008} and \citet{Stuart_2013}. Thus, the researcher fits a predictive model for outcomes in the control group and obtains predictions for potential outcomes for units in the treatment group. 

The test statistic thereby upweights informative covariates associated with the outcome and downweights irrelevant covariates unrelated to the outcome. In our baseline test of as-if random, the coefficients from a linear regression of control group outcomes on covariates are used as weights. The test combines the inputs of standard balance tables---differences of covariate means in the treatment and control groups---into a single prognosis-weighted statistic with one associated $p$-value. This omnibus approach thus also addresses indeterminacy and multiple testing problems. 

We further adapt prognosis weighting for covariate balance tests in three ways. First, we extend the fitted value approach to flexible and nonlinear machine learning methods, which to our knowledge existing work on prognostic scores has not done. We use simulations to assess how prognosis weighting can improve power and specificity, relative to standard unweighted tests, and we compare the different forms of prognosis weighting.

Second, we adapt prognosis weighting for regression-discontinuity designs. To test the continuity of average potential outcomes at the threshold determining treatment assignment, researchers may compare differences of prognosis-weighted intercepts above and below the threshold. As with tests of as-if random, this approach prioritizes the most informative covariates and bases assessment on a single prognosis-weighted statistic.

Third, for hypothesis testing, we provide a bootstrap that accounts for the statistical dependence of covariate control group means and the estimated prognosis weights. We also discuss how to leverage equivalence testing with prognosis weighting. All statistical routines are implemented in our forthcoming R package \texttt{pwtest}.\footnote{Package \texttt{pwtest} can be found on https://github.com/[ANONYMIZED]/pwtest. Installation instructions and syntax are in Online Appendix Section 9).}

Our results show that prognosis-weighted tests can achieve substantial gains in power as well as specificity, relative to standard unweighted tests. The key advantage is that downweighting irrelevant noise variables unrelated to potential outcomes can limit both false positives and false negatives---because conclusions are then based on the most informative covariates. 

% The simulations also suggest the importance of attention to possible non-linear imbalances, on which much applied literature has not substantially focused. 

% As we demonstrate, the prognosis of covariates used in balance tests in prominent political science studies using natural experiments and RD designs varies and is often quite low. 

Finally, we discuss many empirical examples of problems (1)-(3) in applied research and discuss possible solutions. Both the extent of imbalance and, especially, the degree of covariate prognosis vary across the studies in our sample. We report $p$-values for prognosis-weighted tests of as-if random and continuity and show graphically how our prognosis-weighted tests project out irrelevant covariates. This helps demonstrate how prognosis weighting can address the problems we highlight. 

The examples show that by properly prioritizing informative covariates, prognostic-weighted tests base conclusions on the variables most predictive of outcomes. In one set of studies, the relatively predictive individual covariates are imbalanced (e.g. \citealt{Samii2013}, \citealt{BLATTMAN_2009}, or \citealt{Bohlken_2018}); here, prognosis weighting can lead rejections of as-if random, whereas unweighted tests do not. In a second set (e.g. \citealt{Novaes_2018}, \citealt{Kim2019}, and \citealt{Boas2011}), there is imbalance only on irrelevant noise covariates, so adjusting for prognosis may increase our confidence in the validity of the natural experiment.  In a final set of studies (e.g.,\citealt{Fouirnaies2014}), there is a mix of observed balances and imbalances on prognostic and non-prognostic variables, and prognosis weighting helps sort out the relative importance of the different covariates.

We also use as a case study the randomness of close elections, an important topic of recent debate (\citealt{caughey_sekhon_2011}, \citealt{eggers_et_al_2015}, \citealt{DeLaCuesta_Imai}, \citealt{hartman_2021}). Existing tests of balance in close elections exhibit the general problems (1)-(3) we identify. Using data from the U.S. House, we illustrate how prognosis-weighted tests synthesize and extend contrasting previous results. Most importantly, we show that covariate prognosis in cross-national tests of the randomness of close elections is very weak: for example, lagged party incumbency has on average no predictive value for incumbency outcomes in cross-national data. Balance tests using this single covariate, as in \citet{eggers_et_al_2015}, therefore have little power to falsify as-if random or continuity. In particular, they are prone to false negatives.

We focus our discussion of empirical examples substantially on natural experiments and discontinuities, in which analysts usually have outcome data available at the time they conduct balance tests. However, the techniques extend naturally to randomized experiments with attrition, imperfect implementation, or other issues that may be detectable with tests based on covariate imbalance.

Methodologically, by incorporating prognosis-weighting into omnibus tests of identification conditions, we contribute to previous research on covariate balance testing. \citet{ImaiEtAl2008} emphasize the problem that failing to reject a null hypothesis is not the same as accepting it: researchers may fail to reject simply because their study is small and underpowered. Our work adds a further dimension to this ``balance test fallacy," because we show that even a large, apparently well-powered test will not validly test key identification conditions if covariates are not prognostic. In addition, by basing our tests on a single summary test statistic, we provide a new way to address multiple testing concerns (\citealt{DeLaCuesta_Imai}) and complement valuable articles on omnibus covariate balance tests (\citealt{HansenBowers2008}; \citealt{CaugheyDafoeSeawright2017};  \citealt{Gagnon-Bartsch_Shem-Tov2019}), which do not however consider covariate prognosis.\footnote{Related research in statistics and epidemiology recommends upweighting tests for ``important” hypotheses---those most plausibly false---in $p$-value combinations; see e.g. \citet{Fisher1935}, \citet{Holm1979}, \citet{BenjaminiHochberg1997}, \citet{KostMcDermott2002}, \citet{Westfall2005}, and \citet{GenoveseEtAl2006}. Our approach gives specific content to which hypotheses are most likely to be false in balance tests by upweighting covariates related to potential outcomes.} 

Our most important contribution, however, is practical. Social science researchers deploying balance tests in natural experiments and related designs do not currently measure or account for the informativeness of covariates. We show how (1) measuring and (2) maximizing joint prognosis, then (3) prioritizing informative covariates that are predictive of outcomes while de-prioritizing irrelevant ones, can improve the usefulness of covariate balance tests. Prognosis-weighted tests offer an improvement on current practice in applied research---which ignores the issue of prognosis entirely---and can lead to more credible conclusions about whether identifying conditions are met. 

In the next section, we use our sample of papers from top political science journals to illustrate the three key problems we highlight. We then discuss in section \ref{why prog matters} why prognosis matters for testing identification conditions. In section \ref{informative}, we describe bootstrap prognosis-weighted tests of as-if random and adapt the approach to test continuity of potential outcomes in regression-discontinuity designs. We also discuss simulation evidence on the tests' power and specificity. In section \ref{practice}, we turn to practical issues, discussing many examples that show the gains from prognosis weighting and developing the case study of close elections. Our conclusions in section \ref{conclude} expand on our recommendations for practice. Technical details and formal arguments are in section \ref{tech} and the Online Appendix.

\section{The problem of weak covariate prognosis: a survey of reported balance tests}
\label{motivation}

To motivate our focus, we study a random sample of 150 articles that use randomized experiments, natural experiments, and RD designs and that were published in three top political science journals (the \textit{APSR}, the \textit{AJPS}, and the \textit{JOP}), stratifying by journal, over the time period 2000-2019.\footnote{For code used in the sampling, see OMITTED.} Overall, 52 percent of  articles present balance tests. 

The survey suggests three features---and, we will argue, problems---of standard balance tests.

\subsection{Prognosis is rarely measured}

Covariate prognosis is rarely considered systematically.  In fact, we found no examples of efforts to measure the prognosis of the covariates used in balance tests. 

This is a critical omission because different covariates vary in their ability to predict potential outcomes. As we discuss next, (a) the set of covariates used in a given study may or may not be jointly prognostic; and (b) within a given study, different individual covariates may have different degrees of association with the outcome variable. 

Moreover, it is not obvious \textit{a priori} whether a given set of covariates is prognostic. For example, methodologists sometimes recommend using the pre-treatment value of the dependent variable as a covariate (Imbens and Rubin 2015: 483-4, \citealt{EggersTunonDafoe2023}, \citealt{CaugheyDafoeSeawright2017}). This may be because it can be highly prognostic of potential outcomes under control. Yet this also might not hold, for example, due to heterogenous temporal trends or other factors. In our case study of close-election designs  later, we show that lagged party incumbency has almost no predictive value for incumbency outcomes in a cross-national data set. Prognosis is thus an empirical question, and it requires formal diagnosis. 

Reporting measures of the association between covariates and outcomes can give readers an indicator of the informativeness of balance tests. As we will show, prognosis powerfully affects the ability to use covariate balance to test key identification conditions meaningfully (sections \ref{why prog matters} and \ref{informative}). However, such measures are essentially never reported in applied work.

\subsection{Joint covariate prognosis varies across studies---and is often weak}
\label{low prog}

The joint prognosis of covariates used in balance tests in fact varies substantially across different studies---and is often quite low. 

\begin{figure}
    \centering
    \includegraphics[width=0.9\textwidth]{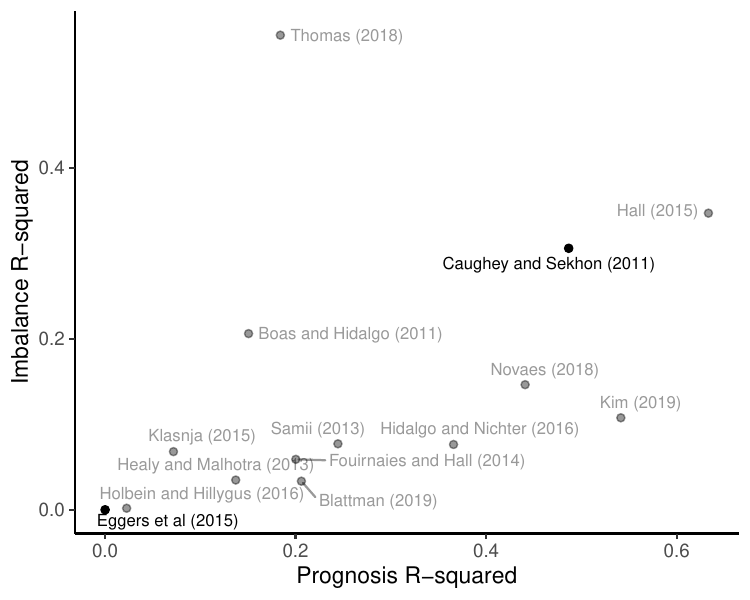}
    \caption{Imbalance vs. Prognosis In Balance Testing (Sample of Natural Experiments and RD Designs)}
   \begin{flushleft}The figure plots a sample of natural experiments and regression-discontinuity (RD) designs drawn from all those published in the \textit{American Political Science Review}, \textit{American Journal of Political Science}, and \textit{Journal of Politics}, 2000-2019; Caughey and Sekhon (2011) is added. Prognosis $R^2$ comes from a regression of potential outcomes under control on all available covariates (control group only). Imbalance $R^2$ comes from a regression of treatment assignment on all available covariates. Two studies we discuss in detail in Section 5 are bolded. See Online Appendix Section 1 for further information.\end{flushleft}
       \label{fig:ImbalancePrognosis}
\end{figure}

The horizontal axis of Figure \ref{fig:ImbalancePrognosis} plots the $R^2$ from the multiple regression of control group outcomes on all available covariates, for a sub-sample of the studies (``Prognosis $R^2$").\footnote{We omitted randomized experiments and stratified on natural experiment versus discontinuity and on the presence of a lagged dependent variable. We excluded some studies due to lack of replication data (Online Appendix Section 6.1).} As we discuss later, such goodness-of-fit measures provide one helpful tool for assessing prognosis. The vertical axis plots the multiple $R^2$ from the regression of a treatment assignment indicator on all available covariates (``Imbalance $R^2$"). 

Figure \ref{fig:ImbalancePrognosis} suggests several insights. First, we find relatively little covariate imbalance in these studies overall. Most cluster along the bottom portion of the plot, with a  low Imbalance $R^2$ (less than 0.1). This likely reflects our sampling strategy: studies with substantial covariate imbalance are unlikely to be published as natural experiments or discontinuity designs. Sampling a fuller range of observational studies would presumably populate the top part of the figure.

Second and more concerningly, however, covariates are not predictive of potential outcomes in many studies.  Some studies located towards the right of the horizontal axis use covariates associated with potential outcomes. Yet, many studies cluster close to the vertical axis---where the prognosis $R^2$ is zero. We note that in the full sample of 150 studies, only 18 percent of balance tests used the pre-treatment value of the dependent variable as a covariate. 

Thus, many balance tests use noise covariates that are only weakly related to potential outcomes. This includes several studies with good observed balance. Arguments we will develop in section \ref{why prog matters} and \ref{informative} suggest that heuristically, there are four kinds of cases in Figure \ref{fig:ImbalancePrognosis}: 
\begin{enumerate}
    \item In the lower-left quadrant, we risk a form of Type I error: we may fail to reject as-if random due to the observed balance of noise covariates unrelated to outcomes. Yet potential outcomes are themselves related to treatment assignment. 
    \item  In the upper-left quadrant, we may instead be prone to spurious rejection of as-if random---because there is imbalance on covariates unrelated to potential outcomes.
    \item In the lower-right quadrant, we find cases with high prognosis but low imbalance: here, the claim of as-if random may be most persuasive. 
    \item Finally, in the upper-right quadrant, rejection may be most persuasive of a failure of as-if random---because imbalanced covariates are as a whole prognostic of potential outcomes. 
\end{enumerate}
For 3 and 4, however, we note that covariates may be associated with potential outcomes as a whole, leading to a high prognosis $R^2$; yet balance or imbalance could occur on a non-prognostic subset of covariates. 

This implies that tests should be based on the most individually prognostic covariates in the set, as in the prognosis-weighted procedures we present in section \ref{informative}.

\subsection{Individual covariate prognosis varies within studies---and is also often weak}
\label{ind variation}

Finally, within studies, different covariates also vary in their informativeness about outcomes.  

\begin{figure}
    \centering
    \includegraphics[width=0.9\textwidth]{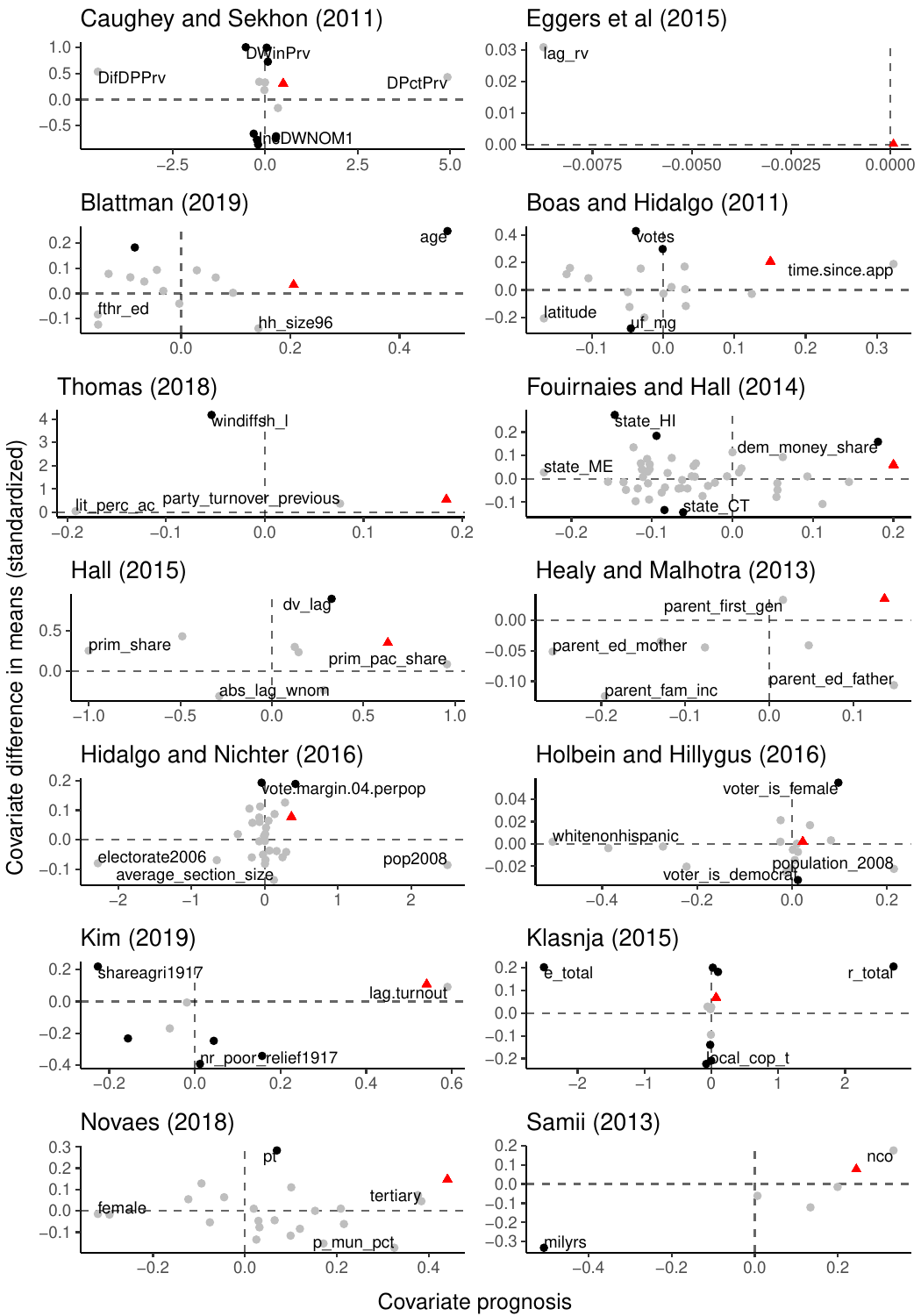}
     \caption{The prognosis and imbalance of individual covariates varies}
     \begin{flushleft}For each of our sampled studies in Figure \ref{fig:ImbalancePrognosis}, we plot for each covariate the standardized difference of means across treatment and control (vertical axis) against the covariate's standardized multiple regression coefficient, from the prognosis regression (horizontal axis). The red triangles indicate the overall prognosis and imbalance $R^2$s. We indicate in black the covariates where the $p$-value $\leq 0.05$ from a two-tailed t-test of covariate values across treatment and control, suggesting covariate imbalance. In discontinuity designs, we use the authors' chosen bandwidths to define the study group for the tests (see Online Appendix Section 1.1.1 for details). \end{flushleft}
    \label{fig:prog_dist}
\end{figure}

In Figure \ref{fig:prog_dist}, we plot (a) the standardized difference of means across treatment and control groups for each covariate in each study in Figure \ref{fig:ImbalancePrognosis} (vertical axis) against (b) that covariate's individual prognosis, i.e., the standardized regression coefficient from the prognosis regression (horizontal axis). The black dots indicate differences of means that are statistically significant in a $t$-test, while gray dots indicate insignificant differences.\footnote{For consistency, we use the same test for each study. The effective sample size can differ across variables due to covariate-specific missing data. See details in Online Appendix Section 1.} 

As Figure \ref{fig:prog_dist} shows, some individual covariates are strongly predictive of potential outcomes---while for many others, the prognosis coefficient is near zero. Furthermore, as with joint prognosis (Figure \ref{fig:ImbalancePrognosis}), the relationship between individual prognosis and the treatment-control imbalances varies. In some studies, irrelevant covariates unrelated to potential outcomes are imbalanced; other times, it is the prognostic covariates that are significantly imbalanced. 

Unfortunately, as with joint prognosis, we find no formal measurement in these studies of which particular covariates are actually predictive of outcomes. Absent a plot like Figure \ref{fig:prog_dist} or other measures of the prognosis of individual covariates, it is difficult to know which ones are informative.

This situation appears typical in the literature. Researchers often present tests for numerous individual covariates: the majority of the studies in our sample that present balance tests (56 percent) report only covariate-by-covariate tests. Such tests can result, however, in indeterminacy (\citealt{KostMcDermott2002}) as well as problems of multiple statistical comparisons (\citealt{BenjaminiHochberg1995}, \citealt{DeLaCuesta_Imai}). Rules of thumb---such as that only 1 out of 20 differences should be significant at the 0.05 level when treatment is randomized---do not apply when covariates are correlated and thus tests are dependent, as they almost always are in practice (\citealt{CaugheyDafoeSeawright2017}). Results of disparate covariate-by-covariate tests fail to lead to a clear decisions rule for rejecting an overall null hypothesis like as-if random.  

Most worrisome, different covariates differ in their informativeness about potential outcomes---yet because prognosis is undiagnosed, it is difficult to know which of the separate covariate-by-covariate tests should be treated as most dispositive. This variation in individual informativeness affects interpretation of test results, as we will show. Even studies that assess the joint imbalance of all covariates using omnibus or global test statistics---for example, by reporting the $p$-value of the $F$-statistic from the (unweighted) regression of a treatment indicator on all covariates---do not account for the varied informativeness of individual covariates.\footnote{By ``omnibus" and "global," we mean a test statistic based on some combination of the covariates that returns a single $p$-value, rather than different $p$-values for different covariates (\citet{CaugheyDafoeSeawright2017}). See \citet{HansenBowers2008} on drawbacks of the $F$-test.} All covariates are thus treated equally in balance tests---but in truth, some are more informative than others.

\subsection{Summary of survey: limitations of standard balance tests}

Our survey shows that covariate prognosis varies both within and across published balance tests. Yet, in none of the studies we examined is prognosis measured. Nor is it incorporated formally into tests. 

What are the implications for our conclusions about identification conditions? How should we think about tests with low imbalance but also low prognosis, as in the bottom-left quadrant of Figure \ref{fig:ImbalancePrognosis}? What inferences can we draw from those located in the figure's other quadrants?  And how should we adapt covariate balance tests to account for the unequal informativeness of different covariates? 

We develop theory and simulations to address these questions in sections \ref{why prog matters} and \ref{informative}, then we return to further discussion of empirical examples in section \ref{practice}.

\clearpage

\section{Are potential outcomes balanced? Why prognosis matters}
\label{why prog matters}

Researchers using experiments, natural experiments, and discontinuity designs often wish to test key identification conditions. 

In valid natural experiments, the following condition must hold:
\begin{assumption}{(As-if Random Assignment)}\label{as:1} \:
\label{POs}
Treatment is assigned independently of potential outcomes.
\end{assumption}
As-if random ensures, for example, that sicker patients do not go systematically to the treatment group in a drug trial studying health outcomes, or that those more prone to vote do not disproportionately receive a vote-mobilizing intervention.\footnote{Assumption \ref{POs} is also sometimes called (strong) ignorability.} If as-if random holds, the true ATE is estimable using simple, transparent methods (\citealt{Freedman1999}).

This assumption can be the ``Achilles Heel" of natural experiments, however (\citealt{Dunning2008}). In a true randomized experiment, a chance protocol under the control of a researcher (\citealt{Fisher1935}) ensures that treatment is independent of potential outcomes, as well as any fixed covariates---though even in experiments, issues like attrition or failed implementation can compromise as-if random. In natural experiments, by contrast, as-if random is held to be an implication of a process, not under the control of the researcher, that is alleged to produce a haphazard allocation that does not depend on potential outcomes. 

Unfortunately, Assumption \ref{POs} cannot be directly verified due to the ``fundamental problem of causal inference" (\citealt{Holland_1986}): we do observe treatment potential outcomes for those assigned to the control group, and vice versa. We thus cannot use the realized distribution of potential outcomes to test as-if random. 

\subsection{What standard balance tests do}
\label{standard}

Researchers therefore seek to test Assumption \ref{POs} using information about the realized distribution of observed \textit{covariates} across treatment and control groups. The difficulty is that these covariates may or may not be related to potential outcomes---and thus standard balance tests therefore may or may not effectively test as-if random.

Suppose that the space of possible covariates contains "signal" covariates, which contain all information about potential outcomes, and ``noise" covariates, which contain none (compare  \citealt{liu2020selfpenalizing}). Signal covariates and potential outcomes are dependent, while noise covariates are independent of potential outcomes.\footnote{See Online Appendix Section 2 for a formal treatment.} 

The logic of standard balance tests appears to rest on the following claim: 

\begin{flushleft} \textbf{Claim 1} Treatment assignment is independent of covariates if and only if treatment assignment is independent of potential outcomes. \end{flushleft} 

\begin{flushleft} Thus, a failed balance test suggests a failure of as-if random, and vice versa.\end{flushleft} 

The claim is false, however. As we show formally in Online Appendix Section 2, two kinds of counterexamples are relevant: 

\begin{flushleft} \textbf{Counterexample to Claim 1: False positives.}  Suppose treatment is assigned independently of potential outcomes, so as-if random holds. However, observed covariates are merely noise, and Nature has adversarially chosen to assign treatment so that it depends on the noise covariates. Then treatment assignment depends on covariates---even though treatment is assigned independently of potential outcomes. The ``if" direction of the claim thus does not follow.\end{flushleft}

For example, in an observational study of the efficacy of a new drug, men might tend to select into the treatment group. Yet gender may be unrelated to health status or responsiveness to the treatment. If we have only data on gender, we may wrongly reject as-if random based on the covariate imbalance, even though potential outcomes themselves may be independent of treatment. 

A researcher who believed Claim 1 might thus perform a balance test, observe imbalance between treatment and control groups on some subset of covariates, and conclude that treatment was not randomly assigned. However, this is a false positive if the imbalanced covariates are unrelated to potential outcomes: their imbalance does not constitute evidence that as-if random fails. 

Conversely---and perhaps most importantly, as we might worry most about false claims to a natural experiment---balance on a spurious covariate does not imply that treatment is assigned independently of potential outcomes, as the next counterexample shows.

\begin{flushleft}\textbf{Counterexample to Claim 1: False negatives.} Assume now that as-if random fails but treatment is assigned independently of the noise covariates. Independence of observed noise covariates and treatment does not therefore imply that treatment is assigned independently of potential outcomes. The ``only if" direction of Claim 1 does not follow.\end{flushleft}

For instance, sicker patients might select into the treatment group. As-if random may thus fail. Health after an intervention may be closely related to prior health---yet we may fail to measure this signal covariate. In contrast, men may be as likely to select into treatment as women, leading to expected balance on gender. Yet, if gender is not related to potential outcomes or responsiveness to treatment, its observed balance cannot readily validate as-if random. If we base a balance test on gender, we may thus falsely fail to reject as-if random.

In sum, covariates differ in their informativeness about potential outcomes. If we only measure noise covariates---those unrelated to potential outcomes---then finding balance or imbalance on those covariates does not allow us to test as-if random assignment. 

\subsection{The power of prognosis}
\label{prog R2}

The discussion thus far suggests we should consider the informativeness of covariates when constructing balance tests.

There are at least two reasons that prognosis of covariates matters for testing---and also thus why covariates with differing degrees of prognosis should not be ``treated equal." 

First, as mentioned, the most direct test of as-if random would assess balance of \textit{potential outcomes} across the treatment and control groups (\citealt{imbens_rubin_2015}, Chapter 21). This test is impossible: once treatment has occurred, we do not observe potential outcomes under control in a treatment group or potential outcomes under treatment in a control group (\citealt{Holland_1986}). 

Yet, a covariate strongly associated with potential outcomes may give us substantial information about this realized balance. Indeed, as we discuss next, if the covariates at our disposal happened to contain all information about subjects' potential outcomes, then we could use the observed balance of covariates to validly test the independence of treatment assignment and potential outcomes. 

Second and relatedly, if subjects self-select into treatment groups, as in many observational studies, then (contra as-if random) the intervention they receive may depend on the outcomes they would experience in each group (\citealt{Heckman1979}, \citealt{AngristPischke2009}). Agents may have unobserved prognostic information that researchers lack, e.g. about their expected gains from treatment that may lead them select into treatment. The informativeness of balance tests can therefore be especially limited if covariates are not predictive of potential outcomes. 

In contrast, as we argue next, measuring and prioritizing prognostic covariates in tests is most likely to detect such selection into treatment on the basis of potential outcomes.  

\subsubsection{A conceptual motivation: minimally sufficient covariates}

As a conceptual motivation for this argument, suppose first that observed covariates that are "minimally sufficient"---essentially, that contain all \textit{and only} the possibly observable information about potential outcomes (\citealt{Dawid_1979}, \citealt{Pearl1988}, \citealt{VanderWeele_Shpitser_2011, VanderWeele_Shpitser_2013}, \citealt{Wang_Wang2020}). In this case, as we show in the appendix, treatment assignment depends on the covariates if and only if they depend on potential outcomes (Online Appendix Section 3.1, Theorem A.1). Then, we may validly reject as-if random based on the non-independence of treatment assignment and covariates---as in standard balance tests.

This idea is useful because it underscores the importance of choosing jointly informative covariates---and then prioritizing the most informative subset for tests. Sufficiency only guarantees the "if" direction of the theorem. Thus, it controls false negatives: if covariates are sufficient, then when treatment is not assigned independently of potential outcomes, we should expect a well-powered balance test to fail. For the "only if" direction, we need minimum sufficiency---i.e., the \textit{and only} part of its definition. This controls false positives: if covariates are minimally sufficient, a failed balance test implies a failure of as-if random. This is a key motivation for the prognosis-weighted tests we discuss in Section 4. By projecting potential outcomes onto covariates before running tests, we effectively discard uninformative covariates---as in the creation of a minimally sufficient from a merely sufficient covariate set. 

In applications, unfortunately, covariates usually cannot be expected to contain such complete information about the values of potential outcomes. Moreover, sufficiency is difficult or impossible to validate. To be sure, a version of this condition could occur in some settings. For example, the pre-treatment value of the response variable may sometimes equal the post-treatment value in the absence of an intervention. Then this covariate's correlation with potential outcomes under control is 1, which also implies sufficiency. Yet, even with a lagged dependent variable, temporal trends could imply differences in values of the outcome variable in the pre- and post-treatment periods, absent an intervention. These trends could be heterogeneous for different units, which implies a correlation less than 1. The direct practical implications of the minimal sufficiency theorem are thus limited.

\subsubsection{The importance of prognosis}

The more prognostic covariates are, however, the closer they may get to approximating the ideal situation of complete informativeness. The logic suggests that the power and specificity of tests will improve as covariates become more predictive of potential outcomes. 

In addition, even when a set of measured covariates is not sufficient, we may be able to improve the performance of tests by prioritizing the individual covariates that are most closely associated with potential outcomes. We discuss simulation evidence that supports this claim in subsection \ref{sims}. 

This argument parallels in some ways standard arguments about the role of prognosis in achieving control over confounding variables in observational studies (\citealt{VanderWeele_Shpitser_2011, VanderWeele_Shpitser_2013}): the more predictive of outcomes the measured covariates are, the less likely it is that unobserved factors produce violations of as-if random. The difference is that here we are focusing on covariate characteristics that allow us effectively to test as-if random, rather than control for violations of it. Testing and estimation can be complementary tasks. However, as we suggested in section \ref{standard}, with balance testing we should be attentive to both false negatives and false positives: we may fail to reject as-if random because spurious covariates are balanced, but we may also falsely reject it because irrelevant covariates are imbalanced. 

The challenge is therefore to collect covariate data that allow us to test identification conditions convincingly. Unfortunately, the predictiveness of covariates for outcomes has been ignored in applications of balance testing in experiments, natural experiments, and discontinuity designs. As we emphasized in Section 2, researchers do not typically report measures of covariate prognosis, nor do they discuss efforts to maximize the informativeness of covariates used in balance tests. Thus, it is usually difficult to assess the extent to which these pathologies of standard balance tests apply. 

In sum, we argue that researchers should attempt to gather the most informative set of covariates possible, and then prioritize the most prognostic ones in their tests. The more prognostic are covariates, the more information they give us about the likely balance of potential outcomes, and the more useful are covariate balance tests. Measures such as the prognosis $R^2$ (Figures 1-2) can be viewed as a continuous operationalization of informativeness. Such measures can help researchers assess the adequacy of information about potential outcomes contained in a set of measured covariates. We return to this point in connection with our simulations (section \ref{sims}), where we assess how error rates in tests vary as a function of levels of prognosis observed in the empirical studies in Figures \ref{fig:ImbalancePrognosis} and \ref{fig:prog_dist}.

\section{Prognosis-weighted covariate balance tests}
\label{informative}

Suppose researchers successfully gather data on jointly prognostic covariates, as recommended in the previous section. How should they combine the information from different covariates to test key identification conditions? 

Prognosis-weighted tests provide a useful approach that prioritizes the covariates that are most informative about potential outcomes. \citet{Hansen_2008} proposed balancing on ``prognostic scores" in observational studies, and some subsequent literature has explored the performance of prognostic-score balancing empirically (see especially \citealt{Stuart_2013}). 

In the technical appendix (section \ref{tech}) and online supplementary materials, we extend these methods to develop covariate balance tests appropriate for assessing different identification conditions. The key test statistic is the difference in the average covariate-adjusted potential outcome under control---that is, a difference of fitted values---across the treatment and control groups. 

In a baseline test of as-if random based on linear regression, this statistic is equivalent to a weighted combination of differences of covariate means, where the weights are coefficients from the prognosis regression---i.e., the multivariate regression of outcomes on covariates in the control group (equation \ref{weight_Delta} in section \ref{tech}). Thus, the test takes the standard inputs of covariate balance tests---covariate differences of means across treatment and control---and combines them into a single test statistic.  

This approach therefore downweights irrelevant covariates and prioritizes informative variables. The goal can also be viewed as constructing a test set as close to minimally sufficient as possible, via a projection of potential outcomes onto covariates before running tests. 

This upweighting of covariates related to outcomes has important consequences for our ability to test identification conditions. In work on prognostic scores, researchers have explored the consequences of omitting prognostic covariates for bias in the estimation of treatment effects (e.g., \citealt{Stuart_2013})---but not, it appears, the consequences of including irrelevant covariates. In the context of covariate balance tests, however, downweighting noise covariates unrelated to potential outcomes can avoid both false positives---rejecting as-if random when it is true---and false negatives---failing to reject as-if random when it is false (see subsection \ref{sims}). 

\subsection{Extensions}
We propose three types of extensions of this basic approach, which are discussed in more detail in section \ref{tech} and the Online Appendix.
\subsubsection{Prognosis weighting based on flexible regressions and non-linear methods}
First, we explore a range of prognosis-weighted tests based on flexible regressions and nonlinear methods, including machine-learning techniques.  

The fitted-value approach leads naturally to these extensions, since different methods can be used to fit covariate-adjusted potential outcomes in the control group and then extrapolate predicted values to the treatment group. The motivation is that by allowing better prediction of potential outcomes---when, for instance, outcomes are nonlinearly related to covariates---flexible methods may improve the power or specificity of tests. 

However, the performance of tests based on machine learning and other flexible methods appears not to have been assessed in the literature on prognostic scores (as noted by \citealt{Stuart_2013} and \citealt{Leacy_2014}). We therefore use simulations to compare the test performance of these different methods (see subsection \ref{sims}). We implement these methods---including linear regression with expanded polynomial bases and covariate interactions, regularization with lasso, and machine learning methods including random forests and gradient boosted trees---as options in our \texttt{R} software (see subsection \ref{flexible}). 

\subsubsection{Prognosis-weighted tests of continuity in regression-discontinuity designs}
Second, we extend the fitted value approach to allow for prognosis-weighted tests of continuity in regression-discontinuity (RD) designs. 

Here, rather than a prognosis-weighted difference of means as in the test of as-if random, the test statistic is a prognosis-weighted difference of intercepts from regressions above and below the threshold determining treatment assignment. 

Conceptually, it is as if we ran separate regressions of each covariate on the running variable in the RD design above and below the assignment threshold, as in standard covariate-by-covariate tests---but then combined the intercepts from the separate regressions using prognosis weighting. 

As with tests of as-if random, the prognosis-weighted difference of intercepts allows a test of continuity based on the covariates that are most informative about potential outcomes (see subsection \ref{rd} and Online Appendix Section 5).  

\subsubsection{Bootstrapped hypothesis tests and equivalence testing}

Finally, we develop statistical hypothesis tests appropriate for these varied forms of prognosis weighting. We discuss a bootstrapping approach that allows naturally for the statistical dependence of the prognosis weights and control group means and that can be readily adapted for clustered or blocked designs, as well as linear and flexible non-linear fitting procedures (section \ref{perm}). 

This bootstrap can be used in connection with the different approaches to fitting covariate-adjusted potential outcomes, including the flexible regressions and machine learning methods. In addition, researchers may use either traditional hypothesis testing or prognosis-weighted equivalence tests to assess identification conditions (section \ref{equiv}). 

Overall, then, prognosis weighting allows covariate balance tests in which rejections of null hypotheses stem from differences of covariate distributions across the treatment and control groups---as in standard approaches---but in which the differences are weighted by measures of prognosis. Since tests may be based on a single prognosis-weighted test statistic and thus one $p$-value, the approach also avoids the problems of indeterminacy and multiple comparisons that beset standard covariate-by-covariate tests. 

% We describe the test statistics and hypothesis tests formally in section \ref{tech} and Online Appendix Sections 4-6. All the techniques are implemented in the forthcoming \texttt{R} package \texttt{pwtest}.

% Under as-if random, the treatment and control groups can be viewed as two simple random samples from the same finite population. Thus, while we cannot observe potential outcomes under control in the treatment group, we would expect a given fitting procedure to lead to the same expected values for fitted values and prognosis weights in both groups, if we could. Thus, in this approach, the researcher fits a model for outcomes in the control group and extrapolates fitted values for units in the treatment group. 

\subsection{Performance of prognosis-weighted tests: evidence from simulations}
\label{sims}

% When covariates are sufficient for $Y(0)$---a best-case scenario---rejecting $H_0$ in a test based on $\delta_{PW}$ implies rejecting as-if random (Appendix Section 2.4, Theorem A.3). However,

Under what conditions does prognosis weighting address the problems we raised in sections \ref{motivation} and \ref{why prog matters}?  
% How does the performance of the tests depend on the joint prognosis of the covariates, and when does prognosis weighting improve on standard unweighted covariate balance tests?
The extent to which prognosis weighting boosts the power and specificity of tests may vary across different data sets and data-generating processes, especially when covariates are not sufficient. This makes the tests' performance well-suited for investigation via simulations. 

We conducted two types of simulations to assess the performance of prognosis-weighted tests. Due to space limitations, we present full results in Online Appendix Section 7.

\subsubsection{Prognosis-weighted vs. unweighted tests}
\label{prog vs unweight}

In one set of simulations, we compare the performance of unweighted to prognosis-weighted tests, while varying the informativeness of observed covariates about potential outcomes (Online Appendix Sections 7.1-7.4). 

Specifically, we compare the prognosis-weighted test using linear regression (subsection \ref{regress}) to two unweighted tests: (i) the sum of standardized covariate differences of means (call this statistic $\delta_{UW}$, see Online Appendix Section 4.2.1) and (ii) Hotelling's $T^2$, another common multivariate test statistic. The latter differ from covariate-by-covariate tests in that they are based on omnibus statistics but are similar to standard approaches in that they treat all covariates ``equally." 

These simulations allow us to study how joint prognosis affects the power and specificity of covariate balance tests and how prioritizing informative covariates, through prognosis weighting, affects performance. 

Thus, we study rejection rates of the tests when as-if random holds and when it is false, varying covariate prognosis (Online Appendix Sections 7.1-7.4).  When as-if random is false, the rejection rate measures statistical power of the test; when it is true, the rate measures false positives (or Type I error, inversely related to specificity).  We consider settings in which observed covariates are sufficient and those in which they are not. In the latter case, we vary prognosis so that covariates are completely uninformative about potential outcomes or only partially so.

\begin{flushleft}\textit{Results: Prognosis-weighted tests avoid both false negatives and false positives}\end{flushleft}

The results illustrate that by projecting out irrelevant covariates, prognosis weighting can reduce both false negatives and false positives. In contrast, unweighted multivariate tests that do not use information on covariate prognosis sacrifice power and/or specificity. The extent to which prognosis weighting improves performance depends on the prognosis of the covariates: as it increases, prognosis-weighted tests become both more powerful and more specific.

Three further conclusions from this first set of simulations are important to note. First, when covariates are sufficient but not minimally so, unweighted tests tend to reject as-if random when it is true or fail to reject it when it is false, due to the balance or imbalance of spurious covariates (Online Appendix Figure A1 and Table A3). In contrast, prognosis-weighted tests control Type I error at standard levels, failing to reject as-if random when it is true; yet it also increasingly rejects as-if random when it is false as the prognosis of imbalanced covariates grows. Thus, compared to unweighted tests, prognosis weighting balances power and specificity: it better detects true failures of as-if random while simultaneously limiting spurious rejections (Appendix Figure A2). 

Second and by contrast, when covariates are insufficient and fully non-prognostic---that is, composed only of noise---weighted and unweighted tests alike are prone to substantial error. When the spurious covariates are balanced in expectation, but as-if random is false, the false negative rate for both kinds of tests approaches 1 (top-left panel of Figure A3). 

Third and finally, however, even when covariates are not sufficient, the power of the weighted---but not the unweighted---tests grows as prognosis increases (right panels of Figure A3 and Table A4). The performance of the prognosis-weighted tests improves as the joint informativeness of measured covariates grows because it prioritizes those individual covariates that are most informative. 

Thus, the simulations illustrate the usefulness of prognosis weighting but also offer an important caveat, consistent with discussion in section \ref{why prog matters}: the quality of balance tests---including prognosis-weighted ones---depends on the overall joint prognosis of measured covariates. Even when covariates are not jointly sufficient, however, the prognosis-weighted test achieves power on the order of 70-80\% in these simulations when the prognosis $R^2$ lies between 0.1 and 0.2 (see Figures A3 and A4 and Table A4). Indeed, prognosis-weighted tests can attain 80\% power at low levels of expected imbalance with prognosis $R^2$ as low as 0.125 (Figure A4). Since these thresholds certainly depend on the data-generating structure, to be conservative, researchers might require a prognosis $R^2$ of 0.2 to defend their tests as meaningfully prognostic. 

The results therefore underscore that diagnosing and reporting covariate prognosis is critical---and so is incorporating information about the relative prognosis of different covariates into tests. 

\subsubsection{Varieties of prognosis weighting: linear vs. machine learning methods}

In a second set of simulations, we compare different types of prognosis-weighted tests.  Here, the prognosis weights are fit using both linear and flexible non-linear methods, under different degrees and types of covariate prognosis and imbalance. We modify the data-generating processes so that potential outcomes are nonlinear functions of the covariates (Online Appendix Section 7.5). 

We first consider simulations with polynomials of the covariates in the process for potential outcomes (Online Appendix subsection 7.5.1). Next, we consider simulations with interaction terms in the outcome process (Online Appendix subsection 7.5.2). Finally, we evaluate prognosis-weighting tests with two `difficult,' highly nonlinear relationships between covariates and potential outcomes (Online Appendix subsection 7.5.3). Thus, we use (a) a ``tree'' specification that creates regime-dependent relationships based on the sign of an interaction term. This specification allows assessment of the methods' performance when the functional form switches discretely based on the interaction term, creating fundamentally different covariate-outcome relationships across regions of the space. We also assess results using (b) a ``sine'' specification incorporates high-frequency nonlinearities. This formulation challenges linear prognosis models with oscillatory components that standard polynomial approximations may not, \textit{a priori}, seem to capture well.  

In these final simulations, we compare the performance of (1) unweighted tests; (2) prognosis-weighted tests based on simple linear regressions; (3) prognosis-weighted tests based on regressions with expanded polynomial bases or interactions; and (4) prognosis-weighted tests based on two machine learning methods, (i) tuned random forests and (ii) tuned gradient boosted trees, as well as lasso with expanded bases (e.g. polynomial and covariate interactions).  We also assess the performance of the best-fitting method (the one that produces the best fit to control potential outcomes) that is automatically selected by the software in each run of the simulations.  We manipulate the target $R^2$ of the prognosis regressions to assess how test performance varies as covariate informativeness changes.

\begin{flushleft}\textit{Results: (Saturated) linear models perform well}\end{flushleft}

The results suggest several useful insights and conclusions. \begin{enumerate}
    \item Prognosis-weighted based on "saturated" linear regressions---i.e., those with expanded polynomial bases or covariate interactions---can sometimes offer improvements in power over simple linear methods, as can flexible nonlinear methods (Online Appendix Figures A5 and A7-A10).
    \item In many simulations, however, the differences are minor. The extent of the improvement depends not just on non-linearities in the relationship between covariates and potential outcomes but also on the nature of imbalances.  When ``main" terms are balanced in expectation but nonlinear (e.g. polynomial or interaction) terms are imbalanced, methods that allow nonlinear fits can offer improved power.  Yet, when there is also expected imbalance on main terms, the expected performance of the tests is often indistinguishable (Online Appendix Figures A7-A10). 
    \item In these simulations, automatic selection of the method that produces the best fit of potential outcomes given covariates in the control group need not lead to the most powerful test. Even with the complex, `difficult' data-generating processes, the expanded linear model with polynomial bases and covariate interactions has the greatest power (Online Appendix Figures A11-A12).  
\end{enumerate}

Overall, the best performer is often the test based on the expanded linear model with polynomials and interactions: it controls Type I error at similar levels as other prognosis-weighted tests when as-if random is true, but it rejects as-if random with the highest probability when it is false. We therefore recommend simple tests with weights based on linear regression, particular with expanded bases where possible, due to their i) estimation stability and sometimes greater power; and ii) the ready interpretability of weights in terms of the relative prognosis of the different covariates.

% Results are broadly similar with the ``sine" process for potential outcomes (Online Appendix Figure A12). Here, too, the saturated linear model has the greatest power, though automatic selection of the best-fitting method also performs well.

% These results suggest that prognosis weighting based on linear methods, particularly those including polynomials of the covariates or covariate interactions, typically performs as well or better than other methods. It is possible that overfitting of outcomes in the control group reduces the power of the machine learning tests and limits gains offered by automatic selection of the best-fitting method. However, 

Our simulations also draw attention to the importance of patterns of linear versus non-linear prognostic imbalances, which to our knowledge has not received attention in work on covariate balance testing. Standard approaches typically test only for main differences (e.g. differences of means), though some do consider differences in distributions using e.g. K-S tests. Analysts should consider the substantive domain under consideration and be attentive to the possibility of non-linear imbalances on prognostic variables. We return to further recommendations in section \ref{conclude}.

% Further work should consider comparisons of methods further, for instance, using more simulations with greater numbers of covariates. 

% We return to further recommendations in section \ref{conclude}.

\section{Prognosis weighting in practice}
\label{practice}

We now return to the three empirical problems identified in the introduction: in applications, (1) covariate prognosis is virtually never measured; (2) joint prognosis of covariates varies across studies and is often low; and (3) prognosis of individual covariates varies within studies and is also often low.  

Our theory and simulations discussed in sections \ref{why prog matters} and \ref{informative} (and the details in section \ref{tech} and the Online Appendix) suggest why these are problems: low-prognosis tests are prone to false negatives and false positives. Without measuring or accounting for prognosis, it is impossible to assess whether conclusions are based on signal or noise covariates. 

Prognosis weighting, by basing conclusions on the most prognostic variables, can help to mitigate problem (3) and leads naturally to diagnostic measures that address problem (1). However, the results of balance tests in studies with low joint covariate prognosis, as in (2), are unreliable. 

Table 1 reports prognosis-weighted and unweighted tests of as-if random for the full sample of studies in Figures \ref{fig:ImbalancePrognosis} and \ref{fig:prog_dist}. The unweighted tests are based on the statistic defined in subsection \ref{prog vs unweight} (see also Online Appendix Section 4.2.1). For studies based on regression discontinuities, we include tests for continuity of potential outcomes. Studies are ordered by the prognosis $R^2$ in the final column, from highest to lowest. We fail to reject as-if random (or continuity, for RD studies) using any test in six of these 14 these papers. In other papers, however, a prognosis-weighted test rejects where an unweighted test does not, or vice versa. 

\begin{comment}
\begin{figure}[ht]\centering\includegraphics[width=0.9\textwidth]{img/fig_studies_delta_pvalue_grouped_paramres.pdf}
     \caption{Prognosis weighting in practice}
     \begin{flushleft}{The figure plots $p$-values for prognosis-weighted (triangles) and unweighted (circles) tests, for the sample of studies in Figure \ref{fig:ImbalancePrognosis}. The number in [brackets] is the prognosis $R^2$. For RD studies, we also include $p$-values from tests of continuity of potential outcomes (gray points). The dotted vertical line is at $p=0.05$. \end{flushleft}
       \label{fig:papersPvalues}
\end{figure} 
\end{comment}

\input{tables/tab_studies_pvalue_summary}

Why do these divergences occur? Inspection of Figure \ref{fig:prog_dist} and comparison to Table 1 suggests that when the weighted test rejects but not the unweighted test, non-prognostic covariates are statistically balanced---but prognostic covariates are imbalanced. 

When the opposite occurs---prognostic covariates are statistically balanced while imbalance occurs on noise covariates---unweighted tests reject while the prognosis-weighted tests do not. 

Finally, a third set of studies in Figure \ref{fig:prog_dist} shows a mix of balance and imbalance on prognostic variables. In this case, prognosis weighting sorts through the mix to allow an overall conclusion. 

In sum, the prognosis-weighted tests appropriately prioritize the most informative covariates. 

\subsection{Three cases where prognosis weighting helps}

We now discuss in more detail examples of these three empirical patterns.\footnote{Details on the analyses for each study are in Online Appendix Section 1.}

\subsubsection{Prognostic imbalance}

In one set of studies, we find balance on noise covariates but greater imbalance on prognostic covariates (see Figure \ref{fig:prog_dist}, Table 1, and Online Appendix Section 1).  In such contexts, an analysis that does not consider covariate prognosis may support the existence of a natural experiment or valid discontinuity design. Yet, prognosis weighting may suggest greater cause for concern.

Consider the excellent study by \citet{Samii2013}, who assesses the consequences for ethnic tolerance of serving in an integrated military in Burundi in the aftermath of a brutal, ethnically charged civil war. Using for identification a discontinuity design based on a military retirement age, the paper shows that serving in an ethnically integrated military decreased prejudicial behavior, though not necessarily the salience of ethnicity. To assess identification conditions, \citet[569-70]{Samii2013} uses balance tests with pre-treatment covariates that have "strong potential to confound were they to exhibit discontinuities near the cutoff." These include noncommissioned officer status; years in the military; years of prewar education; wartime death rates per military unit; and family wartime mortality. Informally, Samii describes why each of these pre-treatment covariate might be linked to ethnic tolerance (the outcome variable). The prognosis $R^2$ is $.194$ (Table 1), indicating that measured covariates are jointly fairly predictive of the outcome. However, as with other standard balance tests in the literature, there is no formal measurement of the informativeness of individual covariates.  

In fact, it turns out that prognosis varies substantially across the individual covariates included in the balance tests. In the standardized prognosis regression of prejudice on covariates, the coefficient is just 0.007 on years of prewar education, while the standardized coefficient on years in the military is -0.468 (see our Figure \ref{fig:prog_dist} and Online Appendix Section 1). Per Figure \ref{fig:prog_dist}, the latter, prognostic covariate is also the only one for which as-if random would be rejected in a covariate-by-covariate difference of means test ($p$-value 0.013). And because the most prognostic covariate is imbalanced, while less prognostic covariates are statistically balanced, the prognostic-weighted omnibus test rejects as-if random ($p$-value 0.018), while an unweighted test does not (Table 1).

\citet{Bohlken_2018} and \citet{BLATTMAN_2009} provide similar examples of imbalance on prognostic variables (Figure \ref{fig:prog_dist}). As shown in Table 1, prognosis-weighted tests therefore call into question as-if random (p-value 0.016 in Thomas; 0.042 in Blattman) and continuity (p-value 0.000 in Thomas). In the case of Blattman's study of the effects of child soldiering in Uganda, the author spotted the imbalance on the prognostic age variable (see Figure \ref{fig:prog_dist}) and controlled for this single covariate in treatment effect regressions. \citet[232]{BLATTMAN_2009} argued that in the Ugandan Lord's Resistance Army, ``abduction parties were under instruction to release only young children and older adults" but other indiscriminately kidnapped "adolescent and young adult males,"  leading to the imbalance on age across abducted (treatment group) and non-abducted (control group) youths.

Formal diagnosis of prognosis---and graphical assessments like those in Figure \ref{fig:prog_dist}---can generally aid the identification of such prognostic, imbalanced covariates. In some studies, like Blattman's, qualitative understanding of the treatment assignment process may support an argument that treatment assignment is independent of potential outcomes---but only conditional on an imbalanced, prognostic covariate. In other studies, identification of such covariates may call into question the assumption of a valid natural experiment or discontinuity design.

\subsubsection{Prognostic balance}

In another set of studies, we find the opposite situation: noise covariates are imbalanced, but there is statistical balance on prognostic variables. In these cases, a naive unweighted test may imply a failed design, but a prognosis-weighted test instead supports key identification conditions.

\citet{Novaes_2018}, for example, uses a close-election design to assess whether barely winning an election causes mayors (who often play the role of partisan brokers) to switch parties at lower rates than near losers.\footnote{\citet{Novaes_2018} also studies whether this effect was influenced by a sudden court decision that restricted elected politicians from switching parties during their term.} The author conducts balance tests on 27 covariates (see our Figure \ref{fig:prog_dist} and Online Appendix Section 1). Our analysis suggests statistical imbalance on several of these variables, but these covariates are all essentially non-prognostic. In contrast, more informative variables are tightly balanced. The $p$-values from prognosis-weighted tests therefore do not reject as-if random or continuity. 

\citet{Kim2019} is another example of a study with substantial imbalance on non-prognostic variables. The author exploits a discontinuity design based on a population threshold that assigns direct democracy to municipalities in Sweden to study effects on the political inclusion of newly enfranchised women. The author tests for balance on nine covariates. Four are negligibly prognostic, with standardized coefficients in the regression of women's turnout (the primary outcome) on covariates that are near zero; two of these noise variables are significantly imbalanced.  By contrast, lagged turnout is highly prognostic, with a coefficient of 0.575 yet also statistical balanced ($p$-value 0.244). Finally, three variables---the pre-treatment tax base, share of agriculture in the economy, and land area---are both moderately prognostic and imbalanced (with respective prognosis coefficients and $p$-values of 0.115 and 0.000; -0.224 and 0.041; and -0.157 and 0.003).  By combining information on imbalance and prognosis across covariates, an informativeness-weighted test helps to sort out these contrasting signals: an unweighted test of as-if random rejects ($p$-value < 0.000) while prognostic-weighted tests reject neither as-if random nor continuity. This is also a highly prognostic set of covariates overall (prognosis $R^2$ of 0.541, per Table 1).

\citet{Boas2011} provide a final example in which there are imbalances in some placebo tests---but only with non-prognostic covariates. In this study, the authors study the effect of political incumbency on control of the media, using a close-elections design.\footnote{City council candidates who barely won an election had more than twice the probability of approval of a community radio license, compared to those who barely lost.} To assess as-if random and continuity of potential outcomes, \citet[876]{Boas2011} present covariate-by-covariate tests for 18 variables. Difference-of-means tests reject the null for three (our Figure \ref{fig:prog_dist}): an indicator for the state of Minas Gerais ("uf\_mg"), the vote share of president Lula of the Worker's Party in 1998, and the log of the electorate size ("votes"). However, these three variables have small standardized prognosis coefficients (-0.045, -0.001, and -0.038, respectively) in the regression of politician’s control of local media on covariates (see Online Appendix Section 1). Other covariates, including the most prognostic covariates such as latitude (prognosis coefficient -0.167) or time since application for a radio license (prognosis coefficient 0.323), are balanced. Boas and Hidalgo (2011: 875) argue that the pattern of imbalance across the 18 covariates "is approximately what one would expect if incumbency status had been randomly assigned." However, rules of thumb for the number of covariates that should be imbalanced in expectation do not readily apply, given dependence across covariates. 

Pooling information across the different covariates to construct a single omnibus, prognosis-weighted test allows a more conclusive test than covariate-by-covariate comparisons. Thus, with the \citet{Boas2011} data, prognosis-weighted tests reject neither as-if random nor continuity (our Table 1).  The tests properly discount imbalances on non-prognostic variables. 

\subsubsection{Mixed prognostic balance and imbalance}

In a final set of studies, there is imbalance on some prognostic and noise covariates, and balance on others.  In these cases, the prognosis-weighted test is crucial for sorting out the relative weight of the balance and imbalance among prognostic covariates and for providing a single test statistic that can summarize the evidence for or against the identification conditions. 

\citet{Fouirnaies2014} provide one example. Per Figure \ref{fig:prog_dist}, a large number of both prognostic and noise variables are statistically balanced, as indicated by the gray shading.  However, several prognostic covariates are also statistically imbalanced. While unweighted tests of as-if random and continuity reject the nulls, prognosis-weighted tests do not.

In sum, in standard covariate-by-covariate tests, analysis of different variables would lead to different conclusions about identification conditions.  And there is no way in typical practice to assess how meaningful is each test, because measures of prognosis are not provided.  Prognosis-weighted tests instead combine information across covariates and properly base conclusions on the most informative of the variables.

\subsection{Case study: are close elections really random?}
\label{close}

Finally, consider a prominent controversy over the randomness of close elections. Contributions to this recent debate illustrate all three of the problems (1)-(3) with covariate balance tests that we have highlighted. Our reanalysis of these data synthesizes previous results but also leads to new substantive conclusions and suggests the need for additional study and data collection.

In a very close election, which party winds up with a slightly greater vote share at time $t$ may seem quite plausibly as-if random (\citealt{Lee_2008}, \citealt{Lee_Lemieux_2010}).\footnote{Later, we discuss another possible identification condition for close-election designs---the continuity of average potential outcomes at the threshold determining treatment assignment.} If true, this facilitates study of the impact of party incumbency on electoral or other outcomes at time $t+1$. 

Yet, in an important study, \citet{caughey_sekhon_2011} critically appraise this assumption for close U.S. House elections (1942-2008). Presenting a series of covariate differences-of-means tests in a small neighborhood around a 0\% difference in Democrat-Republican vote share, they show statistically significant imbalances in past incumbency, as well as the winning party's past vote share, campaign spending, and measures of candidate quality. This appears to undermine as-if random. 

In an excellent subsequent study, \citet{eggers_et_al_2015} extend the Caughey and Sekhon study to a broad range of majoritarian elections around the world. Observing that lagged party incumbency seems to be the major driver of imbalances in Caughey and Sekhon's data, they compare close election winners and losers only on this covariates. They find balance on past incumbency in every other setting they examine. Thus, they conclude that the observed imbalance in U.S. House elections may reflect special features of that context or is simply be due to chance.\footnote{\citet{DeLaCuesta_Imai}, testing for continuity rather than as-if random and correcting for multiple testing, show weaker treatment-control imbalances in the original U.S. House data than Caughey and Sekhon. See also \citet{hartman_2021}, who analyzes these data using equivalence and traditional tests.}

\subsubsection{The three problems of balance testing in close elections}
Consider now, however, how the three problems of covariate balance testing we have discussed impact the conclusions that can be drawn from this controversy.

(1) First, covariate prognosis is not measured or reported. 

\citet{caughey_sekhon_2011} and \citet{eggers_et_al_2015} emphasize the importance of lagged party incumbency, and variables correlated with it, as key covariates for balance tests. However, they do not empirically assess each covariate's prognosis. Nor do they report measures of the overall associations of covariates used in balance tests with outcomes. 

This is important because, as we show next, covariate prognosis in fact varies substantially both within and across these studies---and not always in the way one might expect a priori. This variation has important implications for interpreting the strength of the balance tests. 

(2) Second, individual covariates are not equally informative---yet this variation is not incorporated in tests. 

As shown in Figure \ref{fig:prog_dist}, both the (a) imbalance and (b) prognosis of individual covariates varies substantially in the \citet{caughey_sekhon_2011} data on the U.S. House. Several covariates are imbalanced, but many are not.  And while covariates as a whole appear jointly informative, with a prognosis $R^2$ of 0.49 (Figure \ref{fig:ImbalancePrognosis}), many individual covariates are not prognostic---with standardized regression coefficients near zero. 

% Two variables that show significant imbalances are also highly prognostic: Democratic vote margin (DifDPPrv) and share (DPctPrv) at $t-1$. Yet there are also imbalances on variables that are not prognostic---while many non-prognostic variables are balanced.

% Note that (a) lagged incumbency and several variables correlated with it are imbalanced across districts with Democratic near winners and near losers. Yet, Caughey and Sekhon find balance on other political and demographic variables---such as whether the state has a Democratic governor or secretary of state; the margin of victory in the presidential race; voter turnout; whether the seat is open; and the percentage of urban, Black, or foreign-born residents. 

Unfortunately, there is no formal procedure that takes into account the unequal informativeness of different covariates.\footnote{\citealt{eggers_et_al_2015}, for example, note the imbalance on measures of lagged incumbency; yet while this variable is prognostic in the U.S. House, it is not on average across other countries and elections, as we show shortly, underscoring the importance of formal assessment of prognosis.} And covariate-by-covariate tests offer no ready way to reconcile the contrasting results: some tests reject and others do not, so one cannot readily infer the overall strength of the evidence for or against as-if random.  

Prognosis-weighted ombnibus tests address these problems. As shown in Table 1, the prognosis-weighted test rejects as-if random in Caughey and Sekhon's (2011) data, but an unweighted test does not. This is likely because, as Figure \ref{fig:prog_dist} suggests, there are large imbalances in several prognostic variables. Consistent with \citet{DeLaCuesta_Imai}, the prognosis-weighted test does not reject the weaker assumption of continuity, even though the test is based on the most prognostic covariates. In each case, the omnibus statistic provides a rejection rule based on formal accounting for the varied prognosis of different covariates, while also addressing multiple testing concerns. 

We note also that conclusions using Caughey and Sekhon's data are sensitive to the treatment of missing data.  Our software implementation, by basing the test of as-if random on the vector product of prognosis coefficients and differences of means as in equation (\ref{weight_Delta}) in section \ref{regress}, allows us to use the full set of data available for each difference of means, as in covariate-by-covariate tests.  However, listwise deletion leads to meaningfully different results, in particular an insignificant test statistic for as-if random.  See discussion of the treatment of missing data in Online Appendix Section 8.2.  

(3) Third and perhaps most importantly, weak covariate prognosis in existing cross-national tests does not allow for informative tests of identification conditions. 

Building on the importance of lagged party incumbency in the U.S. House, \citet{eggers_et_al_2015} in fact test for balance \text{only} on this covariate.\footnote{Eggers et al. (2015: 262-3) argue that (a) the variety of characteristics on which winners and losers of close elections may vary can all be viewed as proxies for (are highly correlated with) incumbency; (b) testing for other covariates introduces multiple testing concerns; and (c) incumbency ``confers electoral benefits in a variety of electoral settings around the world."} This approach is entirely understandable as well as practical: lagged party incumbency is readily available across elections and countries, whereas the availability of other pre-treatment covariates may vary by context. 

Yet, the prognostic value of lagged party incumbency in fact varies across countries and types of elections---and in close elections, it is not in fact prognostic on average. Thus, the correlation between the vote share of the incumbent party at time $t - 1$ and time $t$ is $0.79$ across all countries and election types but varies from a low of $0.09$ in Brazilian mayoral elections to a high of $0.91$ in the German Bundestag (full data set); in close elections (defined by a bandwidth of 0.5, i.e., the margin between the winning and runner-up party is less than 1 percentage point), it varies from a high of $0.32$ in New Zealand's post-war parliament to a low of $-0.16$ is the Canadian House of Commons (1867-1911) (see Tables A1-A2 in Online Appendix Section 1.2).

Most concerningly, the average prognosis is essentially zero across all close elections (Figure \ref{fig:ImbalancePrognosis} and Appendix Table A2).\footnote{Restricting the analysis to close elections may attenuate correlations by truncating the range of variation on incumbent vote share at time $t$; yet this is the relevant subset of the data in which to assess prognosis, since this is the set in which balance tests are typically conducted. Note that prognosis is substantially higher in the post-war U.S. House elections studied by Caughey and Sekhon, with a prognosis $R^2$ of 0.83 in the full data and 0.49 in close elections (Figure \ref{fig:ImbalancePrognosis}).}  For reasons we discussed in sections 3 and 4, this weak covariate prognosis implies that the cross-national balance tests are uninformative about the balance of potential outcomes in close elections. 

\subsubsection{Close elections: methodological and substantive conclusions}

Measurement of prognosis and implementation of prognosis weighting helps to synthesize and explain contrasting previous results in the study of close elections. 

Yet, weak prognosis in a cross-national dataset of close elections runs the risk of misleading general conclusions. With the \citet{eggers_et_al_2015} data, prognosis-weighted tests reject neither as-if random nor continuity (Table 1). Yet, as we have shown, balance tests using irrelevant, non-prognostic covariates cannot validly support or falsify key identification conditions. In particular, they are prone to false negatives.  

The failure to measure and account for prognosis---common to all the studies---therefore considerably weakens the conclusions that can be drawn. A finding of statistical balance on a single non-prognostic covariate, as in \citet{eggers_et_al_2015}, cannot compellingly support the general as-if randomness of close elections cross-nationally.

More generally, the results underscore the critical importance of measuring prognosis formally and incorporating it into analyses.   Prognosis is an empirical question. A priori, it appears natural that lagged party incumbency would be highly correlated with future incumbency. In fact, the correlation is negligible across different countries and types of elections.\footnote{See \citet{Schiumerini2025} on the varied effects of incumbency across national contexts and types of elections.} 

These findings imply that the methodological debate about close elections is far from settled. We do not view our results as yet confirming or contradicting the identification conditions in general. They instead suggest the need to leverage a richer set of prognostic covariates for cross-national tests.

\section{Conclusion and recommendations}
\label{conclude}

Covariate prognosis is a critical consideration for balance testing. We have shown that weak prognosis of covariates can lead to both false negatives and false positives in tests of key identification conditions.  Different covariates vary in their informativeness about potential outcomes. Prioritizing more prognostic covariates can increase the power and specificity of tests.

Unfortunately, covariate prognosis receives little attention in prominent studies. Existing applications do not distinguish between informative and uninformative covariates, nor they assess the overall prognosis of the variables used in tests. They may thus not validly test key identification conditions.

We expand in this concluding section on the recommendations summarized in the introduction. 

(1) \textit{Measure and report prognosis}. As the examples from the close-election debate suggested, empirical assessment of prognosis is critical. Goodness of fit measures, such as the Prognosis $R^2$ from a regression of control potential outcomes on covariates, are useful. High values indicate little residual variation in potential outcomes once we condition on covariates. Related measures such as root-mean-squared error can be used to compare prognosis across different covariate sets and fitting methods.

We also urge researchers to inspect the prognosis of individual covariates and the extent the most predictive variables are balanced or imbalanced. Our software implementation facilitates this by producing plots of imbalance against prognosis for individual covariates, as in our Figure \ref{fig:prog_dist}.

Reporting measures of covariate prognosis can help to address an additional concern, which is that a researcher degrees-of-freedom problem can also hinder balance testing. That is, analysts who claim to have discovered a natural experiment might (intentionally or inadvertently) selectively report, omitting tests for the most prognostic covariates if they suggest failures of as-if random. 

Requiring diagnostics of prognosis can ameliorate this problem as well. If reviewers request measures of prognosis to help them assess the likely strength of balance tests, researchers will have incentives to collect data on the most predictive covariates possible. Researchers will then be ``rewarded" (rather than only ``penalized") for using informative covariates in balance tests. 

(2) \textit{Maximize overall prognosis}. Researchers should seek to collect data on the most jointly prognostic covariates possible for use in their balance tests. There is no single recipe for finding such covariates. However, theoretical and substantive knowledge may suggest what pre-treatment variables are likely to be most closely associated with outcomes in a particular context. Often (though not always), the pre-treatment value of the outcome variable is predictive of potential outcomes. Lagged outcomes should thus be prioritized for data collection, where feasible. 

This emphasis on maximizing prognosis also naturally raises the question: how informative must covariates be to allow valid testing of as-if random? While there is no absolute answer to this question, our theory and simulations suggested how different levels of prognosis lead to different error rates. Our simulations suggest reasonable performance even with Prognosis $R^2$ in the range of 0.1-0.2, though this depends on the specifics of the data-generating process and is only intended as a rough guide for adequate informativeness of covariates. In general, the more predictive covariates are of potential outcomes, the more informative and useful are the covariate balance tests.

(3) \textit{Prioritize informative covariates}. Researchers should then prioritize the most individually informative covariates in balance tests, providing an omnibus $p$-value from a prognosis-weighted procedure in either a traditional or equivalence framework.  

Our analysis also raises the question of the specific prognosis-weighting procedure, for instance, whether to use a linear approach to fit prognosis weights or a more flexible regression or machine learning procedure. Indicators of fit from prognosis models can guide the choice of methods. Overall, however, we find that linear tests do remarkably well at boosting power and specificity, even in the presence of highly nonlinear processes for outcomes.  Given the greater simplicity and interpretability of the weighting procedure, we recommend reporting results from a linear method. Saturated models including polynomials and interactions can sometimes increase power over simple linear methods.

As we urged, researchers can also use covariate-by-covariate tables showing prognosis coefficients for each variable. This allows for better understanding of the particular pattern of observed balance and imbalance across prognostic and non-prognostic variables and thus the factors that drive rejection or non-rejection of as-if random or continuity in a prognosis-weighted omnibus test.

We believe that use of the procedures we recommend will lead to more powerful and specific tests of key identification conditions in applied work. At a minimum, it represents an improvement over current practice, in which covariate prognosis is typically completely ignored.  

Covariate balance testing itself as only one component of assessing identification conditions that facilitate causal inference. Qualitative evidence on the process of treatment assignment is important (\citealt{Dunning2012}). Testing itself is complementary to other objectives, including optimization of observed balance to estimate treatment effects as well as sensitivity analysis (\citealt{Rosenbaum2010}). 

Yet, testing identification conditions by examining the distribution of covariates in treatment and control groups should have an important role in design-based analysis of experiments, natural experiments, and discontinuities. Unfortunately, observed balance can be irrelevant when covariates are not associated with potential outcomes. By instead leveraging the power of prognosis, researchers can build more useful, informative tests.   

\section{Technical appendix}
\label{tech}

In this appendix and in the online supplementary materials, we formalize the arguments in the paper and discuss ancillary results.

\subsection{A prognosis-weighted test statistic}
\label{finite}

Consider first a study with a finite population of $N$ units indexed by $i=1,\ldots,N$ and one treatment and one control condition. Let $Y_i(1)$ and $Y_i(0)$ be potential outcomes under exposure to treatment and to control, respectively. The causal effect for each unit is $\tau_i = Y_i(1) - Y_i(0)$, while the Average Treatment Effect (ATE) is $\tau = \mathbb{E}[Y_i(1) - Y_i(0)]$, where the expectation is taken over the draw of a single unit at random from the finite population.\footnote{This formalization embeds the stable unit treatment value assumption (Cox 1958, Rubin 1978).} The random variable $Z_i \in \{0,1\}$ denotes treatment assignment, with $0$ for the control group and $1$ for the treatment group; an $N \times 1$ random vector $Z$ collects the $Z_i$. This set-up is design-based in that the only source of random variation is the treatment assignment vector $Z$; potential outcomes are fixed.

As-if random (Assumption \ref{POs}) motivates the following testable null and alternative hypotheses:
\begin{eqnarray}
\label{hyp}
 H_0&:&  \mathbb{E}[\overline{Y(0)^{T}} - \overline{Y(0)^{C}}]=0 \hspace{5 mm} \nonumber \\
H_A&:&  \mathbb{E}[\overline{Y(0)^{T}} - \overline{Y(0)^{C}}] \neq 0.
\end{eqnarray}
Here, $\overline{Y(0)^{T}}$ is the average value of potential outcomes under control in the treatment (``T") group sample, while $\overline{Y(0)^{C}}$ is the average value of potential outcomes under control in the control (``C") group sample. Both are random variables when treatment assignment is randomized. 

The logic: if as-if random holds, the treatment and control group averages can be viewed as the means of samples drawn at random from the same finite population. Thus, the expected averages are the same in each sample, as under the null hypothesis $H_0$. Conversely, if treatment assignment were not randomized so that $Z$ depends statistically on \{Y(1), Y(0)\}, it follows that the average potential outcomes in the treatment and control groups differ in expectation---as under $H_A$. 

To test $H_0$, the problem is to estimate the unobserved difference of expectations in (\ref{hyp}). This in turn requires a procedure for predicting $\overline{Y(0)^{T}}$ in the treatment sample, where potential outcomes under control are not observed. Then, we can form a test statistic as the difference
\begin{equation}
\label{y_hat diff}
\delta_{PW} = \widehat{\overline{Y(0)^{T}}} - \widehat{\overline{Y(0)^{C}}},
\end{equation}
that is, the fitted average $Y(0)$ in the treatment group minus the fitted average $Y(0)$ in the control group. Essentially, we fit $\widehat{Y(0)}|X$ in the control group, which gives us prognosis weights, and then apply this weighting procedure to the covariates in the treatment group. 

In sum, the test statistic $\delta_{PW}$ is the prognosis-weighted ("PW") difference ("$\delta$") in fitted values across the treatment and control groups. We focus on fitted potential outcomes under control, as in e.g. \citet{Hansen_2008} and \citet{Stuart_2013}. This is because pre-treatment values of the outcome variable, which are sometimes measured, may tend to be especially prognostic for $Y(0)$. 

\subsection{A regression-based test}
\label{regress}

We focus first on linear regression-based fits for the test statistic in (\ref{y_hat diff}). A test based on this approach leads to a simple and readily interpretable test statistic: the weighted difference of covariate means across the treatment and control groups, where the weights are measures of prognosis.

Consider first the sample regression of the outcome variable on covariates in the control group:
\begin{eqnarray}
\label{sample_ave_control}
\widehat{\overline{Y(0)^{C}}} &=& \overline{X^{C}} \hspace{1.5 mm} \widehat{\beta^{C}} \\
&=& \overline{Y(0)^{C}}, \nonumber 
\end{eqnarray}
where the $1 \times p$ vector $\overline{X^{C}}$ gives the average value of the $p$ covariates in the control group and the $p \times 1$ vector $\widehat{\beta^{C}}$ gives the coefficients from the control group regression.\footnote{\label{beta def}Here, $\widehat{\beta^C}=(\sum_{i=1}^{n_0} X_iX_i^{'})^{-1} \sum_{i=1}^{n_0} X_iY_i(0),$
is a $p \times 1$ vector with elements $\widehat{\beta_j}$ for $j=1,\ldots,p$. Here we index by $i=1,\ldots,n_0$ the random subset units sampled into the control group from the $N$ units in the finite population.} Descriptively, the control group regression evaluated at the average value of the covariates is exactly the sample average $\overline{Y(0)^{C}}$. While we cannot fit the analogous finite-population regression---because we do not see $Y_i(0)$ for units in the treatment group---under as-if random the control group is a simple random sample from the finite population. Equation (\ref{sample_ave_control}) can thus be viewed as a regression-weighted estimator for the average potential outcome under control in the finite population (\citealt{Cochran1977}, Chapter 7). 

As for the treatment group, we cannot run a regression like equation (\ref{sample_ave_control}), because in the treatment sample we observe $Y(1)$ rather than $Y(0)$. However, under a null hypothesis of as-if random, the expectation of the coefficient we would obtain---if we could run the regression in the treatment sample---is clearly the same as the expectation of $\widehat{\beta^{C}}$. We can therefore estimate the average of the potential outcomes under control in the treatment group as
\begin{equation}
\label{sample_ave_treatment}
\widehat{\overline{Y(0)^{T}}} = \overline{X^{T}}\hspace{1.5 mm}  \widehat{\beta^{C}},
\end{equation}
where $\overline{X^{T}}$ is the vector of average values of covariates in the treatment group.

Subtracting (\ref{sample_ave_control}) from (\ref{sample_ave_treatment}) gives an estimator of the unobserved difference of the expectations (\ref{hyp}), valid under $H_0$. Thus we have
\begin{eqnarray}
\label{weight_Delta}
\widehat{\rm E}[\overline{Y(0)^{T}} - \overline{Y(0)^{C}}] &=& (\overline{X^{T}}-\overline{X^{C}})  \hspace{1 mm} \widehat{\beta^{C}} \nonumber \\
&=& \sum_{j=1}^{p} \widehat{\beta_j^{C}}\delta_j \\
&\equiv& \delta_{PWLR}, \nonumber
\end{eqnarray}
with ``PWLR" for "prognosis-weighted linear regression." 

Thus, the key test statistic $\delta_{PWLR}$ is the prognosis-weighted difference of covariate means across the treatment and control groups. Each $\delta_j$ in (\ref{weight_Delta}) is the difference of means on covariate $j$, while the weight $\widehat{\beta_j^{C}}$ is the $j$th coefficient from the control-group regression of outcomes on covariates. To ensure that the contribution of each term to the sum is not a function of the measurement scale, we recommend standardizing $Y(0)$ and all covariates; this is the default option in our \texttt{R} package \texttt{pwtest}. The standardized regression coefficients will be larger in absolute value for more prognostic covariates, while they vanish when the partial correlation between $Y(0)$ and $X_j$ is zero. 

It is important to emphasize that $\beta$, the coefficient of the finite- population regression corresponding to the sample regression in (\ref{sample_ave_control}), has no causal interpretation: the regression simply provides the best linear approximation of the potential outcomes $Y(0)$ given $X$. Covariates are fixed features of units that are not here considered amenable to manipulation; even if they were, there is no expectation or requirement that manipulation would lead to expected changes in the value of the outcome variable. The procedure simply allows for measurement of covariate prognosis. Thus, as with other procedures we consider next, the test statistic $\delta_{PWLR}$ combines information on prognosis across covariates to form an ombnibus statistic to which we may attach a single $p$-value to test $H_0$. This may lead to more powerful and specific tests than do standard procedures (subsection \ref{sims}). 

The use of the regression-based test has several possible advantages, relative to more flexible procedures we consider in subsection \ref{flexible}. One is its simplicity and intelligibility: here, the weights (regression coefficients) are readily interpretable as the (linear) prognosis of the respective covariate, relative also to the other covariates.\footnote{This is by the Frisch–Waugh–Lovell (FWL) theorem or ``regression anatomy" (\citealt{AngristPischke2009}: 3.1.2). Each element $\beta_j$ of $p \times 1$ vector of coefficients $\beta$ in the analogous finite-population regression can be represented as the coefficient from the bivariate regression of $Y(0)$ on the residual of $X_j$ on the other $p-1$ covariates.} Another is its close connection to current practice. The test uses the inputs of standard balance tests---covariate differences of means---and combines them into a single, readily interpretable prognosis-weighted test statistic. Thus, rejection of as-if random in tests using $\delta_{PWLR}$ will be due to treatment-control differences of covariate means, as in standard covariate-by-covariate tests. Yet, unlike standard practice, the test prioritizes the variables most informative about potential outcomes.  We assess the relative performance of different tests in subsection \ref{sims}.

\subsubsection{A resampling-based (bootstrap) hypothesis test}
\label{perm} 

For hypothesis testing, we propose a resampling (a.k.a. bootstrap) technique which allows comparison of the observed value of a test statistic to its exact randomization distribution.\footnote{On randomization tests, see \citet{Fisher1935}; also inter alia \citet{CaugheyDafoeSeawright2017}.} The procedure uses draws from the observed data to approximate the null sampling distribution of $\delta_{PW}$, i.e., its distribution when as-if random holds. Thus, for $\delta_{PWLR}$, we draw two independent samples of potential outcomes from the control group; fit the prognosis regression in one of them; calculate a bootstrap test statistic, i.e., the prognosis-weighted difference of means; and repeat the bootstrap $B$ times in order to compare an observed test statistic to its randomization distribution. For non-linear fitting methods discussed next, the procedure is parallel but uses the chosen non-linear regression or machine learning procedure in place of linear regression. 

The validity of the bootstrap rests on two key features.  First, the expectation of the covariate difference of means e.g. in $\delta_{PWLR}$ is zero, as it is when treatment assignment is randomized. Thus, we compare the expected values of averages of two independent samples drawn from the same finite bootstrap population.\footnote{The observed treatment and control group means are dependent and the samples are drawn without replacement. However, $X_i$ is the same whether unit $i$ is assigned to treatment or control. Per Neyman (1923), it is thus as if the two samples were drawn independently with replacement (see \citealt{FPP2007}: A32-A34; \citealt{SamiiAronow2012}, Theorem 2; \citealt{GerberGreen2012}: 57; or \citealt{Dunning2012}: 193).} Second, the procedure allows in a natural way for the statistical dependence between the random variable $\widehat{\beta}^C$---as realized in the control group---and $\overline{X^C}$, with treatment assignment as the only source of stochastic variation. Note that the bootstrap uses only values of $Y(0)$ from the control group to simulate the distribution of prognosis weights. 

This bootstrap procedure can be adapted to accommodate a wide range of designs, for instance, those with clustered or blocked assignment. We also note that using control group values to estimate the weights does not induce a bias from overfitting, a problem that can arise when study outcomes are also used for estimating average treatment effects (\citealt{Rubin2007, Hansen_2008, Liao_et_al}). Further details are in the Online Appendix (Section 4).

% We give details on the construction of the statistic in Appendix Section 2 and develop a resampling (a.k.a. bootstrap) hypothesis test that is valid both for $\delta_{PWLR}$ and for machine learning tests we discuss next. The procedure uses draws from the observed data to approximate the null sampling distribution of the test statistic, i.e., its distribution when as-if random holds, which allows comparison of the observed value of a test statistic to its exact randomization distribution.\footnote{On randomization tests, see \citet{Fisher1935}; also inter alia \citet{CaugheyDafoeSeawright2017}.} The procedure allows in a natural way for the statistical dependence between the prognosis weights---as estimated in the control group---and the control group average, with treatment assignment as the only source of stochastic variation. 

\subsection{Flexible non-linear tests}
\label{flexible}

The fitted value approach also leads naturally to alternative, more flexible nonlinear techniques. The predicted potential outcomes in $\delta_{PW}$ in equation (\ref{y_hat diff}) can be formed by a host of methods. 

In subsection \ref{sims} and Online Appendix Section 7, we explore the performance of two main alternatives. First, we extend the linear regression-based approach of subsection \ref{regress} to include polynomial terms and a full set of covariate interactions. Thus, a fully ``saturated" regression produces the fitted values.

Second, we extend our software \texttt{pwtest} to allow for a host of more flexible methods, including machine learning (ML) techniques.  The options include, among others, generalized linear models with LASSO, Bayesian Additive Regression Trees (BART), random forests, and gradient boosted trees. The strategy is the same across all methods and follows the following steps: 
\begin{enumerate}
    \item Fit $\widehat{Y(0)^C}$ on covariate set $X^C$ (i.e., subsetting to control units), using a given method;
    \item With the resulting fit, obtain $\widehat{Y^T(0)}$ using treatement-group covariate values $X^T$; and 
    \item Calculate the observed $\delta_{PW}$ as defined by equation (\ref{y_hat diff}). 
\end{enumerate} 
The software bootstraps a hypothesis test and associated $p$-values using the approach described in subsection \ref{perm} and returns diagnostic measures of prognosis. Details are in Online Appendix Section 5.

In subsection \ref{sims}, we use simulations to assess the performance of the saturated regression and two widely used ML methods---gradient boosted trees and random forests (\citealt{Breiman2001, Hastie_2009, Zhou_2012, Chen_2016})---as well as the performance of a procedure for choosing the "best"-fitting model that we discuss next. 

\subsubsection{Cross-validation and choice of methods}

Our \texttt{pwtest} function also allows for an automated selection of the method with the best predictive performance. In this case, the method that predicts $Y(0)$ most accurately from covariates in the control (or training) group is selected for use in the resulting test procedure. For the ML methods, this is also based on a $k$-fold cross-validation process for selection of hyperparameters using control group units only. To select an appropriate fitting procedure in a data-driven way, the software picks the estimation method with the highest $R^2$ on the task $Y(0) | X$ in the control group. We discuss further details in subsections 4.4 and 7.5 of the Appendix.

Ideally, the procedure for selecting the fitting procedure should be pre-specified in advance of testing. However, we also note that for reasons described in subsection \ref{regress}---especially simplicity and interpretability of the weights---there may often be a rationale for using the baseline linear approach (the default in \texttt{pwtest}), even if non-linear methods can provide a slight improvement in power, and we recommend also reporting tests using this simple approach. We return to discussion of this point in connection with the simulation results in subsection \ref{sims}.

\subsection{Regression-discontinuity designs: testing the continuity of potential outcomes}
\label{rd}

We now turn to the adaptation of prognosis-weighting to regression-discontinuity (RD) designs. 

Analysts have rightly noted that in many RD designs, as-if random should be replaced with the (weaker) assumption that the regression functions relating potential outcomes to the forcing variable (a.k.a. the running variable or ``score") are continuous at the threshold determining treatment assignment (\citealt{CalonicoCattaneoTitiunik2014}; \citealt{DeLaCuesta_Imai}).\footnote{This may especially be so when the slope of the regression function relating potential outcomes to the forcing variable is not flat (see \citealt{Dunning2012} Chapters 3 and 5; \citealt{CattaneoEtAl2015a}; \citealt{SekhonTitiunik_RD}). When it is flat, the assumption of as-if random within a small bandwidth around the threshold may be the relevant condition to test, using the techniques discussed in sections \ref{regress} and \ref{flexible} with a few modifications (Appendix Subsection 9.1).}

In this case, the key condition to test is not as-if random (Assumption \ref{POs}) but rather:

\begin{assumption}{(Continuity of Potential Outcomes---RD Designs)}\label{as:1} \:
\label{cont_assumption}
Potential outcomes regression functions are continuous at the threshold determining treatment assignment. 
\end{assumption}
Continuity implies that the limits of the regression functions are the same approaching from above and below the threshold. This motivates the standard approach of testing for the equality of intercepts of two regressions, fit above and below the threshold value of the running covariate.

However, researchers typically test for the continuity not of potential outcomes---but of \textit{covariates}. Thus, they regress each pre-treatment covariate separately on the forcing variable, above and below the RD threshold, and conduct a test for equality of the intercepts at the threshold. 

Unfortunately, such tests for the continuity of covariates may not be informative about the continuity of \textit{potential outcomes}. Just as with tests of as-if random, researchers are subject to false negatives and false positives due to irrelevant covariates (section 3.2). Covariates may be continuous at the threshold and yet potential outcomes may not be; or vice versa.  The standard approach also raises the problems of indeterminacy and multiple testing (\citealt{DeLaCuesta_Imai}), as in covariate-by-covariate tests of as-if random.  

Fortunately, we can readily form a prognosis-weighted test statistic that is appropriate for testing continuity of potential outcomes in RD designs. Following our previous approach of using only the prognostic part of the covariates, we first project the outcome variable on covariates on the control group side of the RD threshold. Then, we fit regressions---not of covariates, as in standard practice, but of fitted potential outcomes---on the running variable, on each side of the threshold.  

Thus, let $\widehat{Y(0)} = \bm{X} \hspace{.5 mm} \widehat{\beta^C}$ be the fitted value from a regression of the outcome on covariates on the control group side of the RD threshold, where $Y(0)$ is observed. Now, following standard presentations of RD estimation methods (see e.g. \citealt{DeLaCuesta_Imai}), we fit two regressions. First,
\begin{equation}
\label{minY0left}
(\widehat{\alpha_0} , \hspace{1 mm} \widehat{\beta_0}) = \underset{\alpha_0, \hspace{.5 mm} \beta_0}{\arg \min} \sum_{i=1}^n \mathbb{I}\{c_0 \leq R_i \leq c\}\{\widehat{Y_i(0)} - \alpha_0 - \beta_0(R_i - c)\}^2 K\left(\dfrac{R_i  - c}{h}\right)
\end{equation} 
is the intercept and slope from a regression of $\widehat{Y(0)}$ on the forcing variable to the right of the threshold (centered at the threshold). Here, $R_i$ is the forcing variable, $c$ is its value at the assignment threshold, and $c_0$ is the value that defines the edge of the control-group bandwidth. Similarly,
\begin{equation}
\label{minY0right}
(\widehat{\alpha_1} , \hspace{1 mm} \widehat{\beta_1}) = \underset{\alpha_1, \hspace{.5 mm} \beta_1}{\arg \min} \sum_{i=1}^n \mathbb{I}\{c< R_i \leq c_1\}\{\widehat{Y_i(0)} - \alpha_1 - \beta_1(R_i - c)\}^2 K\left(\dfrac{R_i  - c}{h}\right)
\end{equation} 
is the intercept and slope from the regression on the treatment-group side, including units up to $c_1$.\footnote{As recommended by \cite{CalonicoCattaneoTitiunik2014} and \cite{cattaneo_idrobo_titiunik_2020}, equations (\ref{minY0left}) and (\ref{minY0right}) are triangular kernel-weighted local linear regressions; $K(\cdot)$ may be a function such as the triangular kernel, $K(u) = (1-|u|) \cdot \mathbb{I}\{|u| < 1\}.$  The bandwidth $[c_0, \hspace{.5 mm}c_1]$ can be chosen by the algorithm of \cite{ImbensKalyarnaraman_optimalbandwidth_2012}; this is the default option in our \texttt{R} package \texttt{pwtest}.}  

Conceptually, it is as if we regressed each pre-treatment covariate on the forcing variable in windows below and above the assignment threshold, as in standard practice. However, we combine these separate regressions into one omnibus prognosis-weighted test statistic,
\begin{equation}
\label{tau rd}
\delta_{PW}^{RD} \equiv \widehat{\alpha_1} - \widehat{\alpha_0},
\end{equation} 
where $\widehat{\alpha_0}$ and $\widehat{\alpha_1}$ are the intercepts at the assignment threshold of the regressions of $\widehat{Y(0)}$ on the forcing variable, on the control-group and treatment-group sides respectively (Online Appendix Section 5.2.2).\footnote{For clarity, we separate the fitted intercepts $\widehat{\alpha_0}$ and $\widehat{\alpha_1}$ from $\widehat{\beta_0}$ and $\widehat{\beta_1}$, the fitted coefficients on the centered value of the forcing variable, $R_i - c$. Note, however, that the latter are distinct from the fitted coefficients of the regression of $Y(0)$ on \textit{covariates} $\bm{X}$. These, which we label $\widehat{\beta^C}$ as before, are fit in the prognosis regression.} We can then test the null hypothesis that the expectation of this difference is zero against the alternative of a non-zero difference (or we can flip the null and alternative, as in equivalence testing, discussed next). As the test is based on a single omnibus statistic, it also avoids the problems of indeterminacy and multiple testing associated with covariate-by-covariate tests.

In sum, the test of continuity---as with the test of as-if random---projects out irrelevant covariates and bases assessment on the most informative covariates. See Online Appendix Section 9.2 for further details. 

\subsection{Equivalence testing}
\label{equiv}

Finally, prognosis weighting can also be adapted to take advantage of equivalence tests (\citealt{HartmanHidalgo2018}). Equivalence tests seek to address the ``balance test fallacy" (\citealt{ImaiEtAl2008}, Section 7), in particular, the problem that failing to reject the null of as-if random is not the same as accepting it. With traditional tests, researchers may fail to reject simply because a study is small and underpowered. 

The test works by switching the null and alternative hypotheses, so that under the null, the expected means in the treatment and control group differ, while under the alternative they are approximately equal. Equivalence tests are less likely to reject the null of difference as study size shrinks (\citealt{HartmanHidalgo2018}, Figure SI-2), so acceptance (rejection of the absence) of as-if random is less likely to be an artifact of low power.

Prognosis-weighted equivalence tests can provide an additional protection against the balance test fallacy. In Online Appendix Section 5, we adapt the bootstrap procedure in subsection \ref{perm} for equivalence testing. Here, the most informative covariates must be sufficiently balanced to reject the null hypothesis of difference. Thus, as long as covariates are sufficiently jointly informative, prognosis weighting ensures that we will not ``accept" as-if random unless covariates related to potential outcomes are sufficiently balanced. 

It is important to note, however, that an equivalence test based on covariates with weak joint prognosis is subject to similar limitations as traditional tests. Thus, we may reject the absence of as-if random based on the balance of the most prognostic individual covariates, among the set at our disposal. Yet, if measured covariates are not as a whole prognostic, there could readily be lurking prognostic variables that are unobserved and imbalanced. Were we successfully to measure these prognostic covariates, we might instead reject (fail to reject the absence of) as-if random.\footnote{A further drawback is that researchers may find evidence for or against as-if random by varying the equivalence range.  Alternatives that lessen this discretion---for instance, use of the equivalence confidence interval (\citealt{HartmanHidalgo2018})---make equivalence testing more akin to traditional balance testing since in the latter, one can also readily examine a $(1-\alpha)*100\%$ confidence interval to see what parameter values lie outside of it.}

The way around this difficulty---as with traditional testing---is to ensure that we have measured covariates that are adequately jointly prognostic. The best advice may be thus to develop high-powered tests---either traditional or equivalence-based---by leveraging jointly prognostic covariates and then prioritizing balance of the most informative individual covariates, as in our prognosis-weighted test.  

\nocite{HALL2015}
\nocite{Kim2019}
\nocite{Hidalgo_Nichter_2016}
\nocite{Thomas_2018}
\nocite{BLATTMAN_2009}
\nocite{Boas2011}
\nocite{Healy2013}
\nocite{Klasnja_2015}
\nocite{Fouirnaies2014}
\nocite{Holbein_Hillygus_2016}

% \printbibliography
\bibliographystyle{apalike}
% \bibliography{refs2,refs3}

\bibliography{refs2.bib, refs3.bib}

%TC:subst \printbibliography \bibliography
%TC:macro \field [0,1]
%TC:macro \name [0,0,0,1]
%TC:macro \list [0,0,1]
%% Use \printbibliography if using BibLaTeX
% \printbibliography

% \bibliography{refs2.bib}

%TC:subst \printbibliography \bibliography
%TC:macro \field [0,1]
%TC:macro \name [0,0,0,1]
%TC:macro \list [0,0,1]
%% Use \printbibliography if using BibLaTeX
% \printbibliography
\end{document}

%% file: tables/tab_studies_pvalue_summary.tex
% latex table generated in R 4.3.1 by xtable 1.8-4 package
% Wed Aug 27 21:05:49 2025
\begin{table}[ht]
\centering
\caption{Summary of p-values from prognosis-weighted and unweighted tests for the sample of studies in Figure 1. For RD studies, we include p-values from tests of continuity of potential outcomes. Studies are ordered by the value of the prognosis $R^2$, from highest to lowest.} 
\label{studies_pvalue_summary}
\begin{tabular}{llrrr}
  \hline
Study & Test & $\delta_{UW}$ $p$-value & $\delta_{PW}$ $p$-value & Prognosis $R^2$ \\ 
  \hline
Hall (2015) & as-if random & 0.166 & 0.988 & 0.633 \\ 
   & continuity & 0.580 & 0.653 &  \\ 
  Kim (2019) & as-if random & 0.000 & 0.792 & 0.541 \\ 
   & continuity & 0.592 & 0.826 &  \\ 
  Caughey and Sekhon (2011) & as-if random & 0.470 & 0.004 & 0.487 \\ 
   & continuity & 0.994 & 0.812 &  \\ 
  Novaes (2018) & as-if random & 0.676 & 0.676 & 0.441 \\ 
   & continuity & 0.644 & 0.090 &  \\ 
  Hidalgo and Nichter (2016) & as-if random & 0.540 & 0.548 & 0.366 \\ 
   & continuity & 0.308 & 0.323 &  \\ 
  Samii (2013) & as-if random & 0.480 & 0.030 & 0.244 \\ 
   & continuity & 0.168 & 0.298 &  \\ 
  Blattman (2019) & as-if random & 0.162 & 0.042 & 0.206 \\ 
  Fouirnaies and Hall (2014) & as-if random & 0.034 & 0.274 & 0.200 \\ 
   & continuity & 0.002 & 0.658 &  \\ 
  Thomas (2018) & as-if random & 0.000 & 0.016 & 0.184 \\ 
   & continuity & 0.000 & 0.000 &  \\ 
  Boas and Hidalgo (2011) & as-if random & 0.122 & 0.312 & 0.151 \\ 
   & continuity & 0.564 & 0.434 &  \\ 
  Healy and Malhotra (2013) & as-if random & 0.216 & 0.538 & 0.137 \\ 
  Klasnja (2015) & as-if random & 0.328 & 0.000 & 0.072 \\ 
   & continuity & 0.860 & 0.685 &  \\ 
  Holbein and Hillygus (2016) & as-if random & 0.310 & 0.012 & 0.023 \\ 
   & continuity & 0.854 & 0.273 &  \\ 
  Eggers et al (2015) & as-if random & 0.464 & 0.666 & 0.000 \\ 
   & continuity & 0.218 & 1.000 &  \\ 
   \hline
\end{tabular}
\end{table}